\documentclass[aps,pra,showpacs,superscriptaddress,groupedaddress, twocolumn,nofootinbib]{revtex4-2}  

\usepackage[utf8]{inputenc}
\usepackage{bm}
\usepackage{color}
\usepackage{hyperref}
\usepackage{amsmath, mathtools}
\usepackage{amssymb}
\usepackage{wasysym}
\usepackage{graphicx} 
\usepackage{booktabs} 
\usepackage{pdfsync}
\usepackage{verbatim}
\usepackage{latexsym}
\usepackage{dsfont}

\usepackage{float}

\def\sec#1{\section{#1} }
\def\ssec#1{\subsection{#1} }

\def\({\left(}
\def\){\right)}
\def\[{\left[}
\def\]{\right]}

\def\a{\alpha}

\def\f#1#2{\frac{#1}{#2}}

\def\d{\partial}
\def\de{\delta}

\def\s{\sigma}

\def\th{\theta}

\def\<{\langle}
\def\>{\rangle}

\usepackage{graphicx}
\usepackage{dcolumn}
\usepackage{bm}

\begin{document}

\title{Relativistic Ritz approach to hydrogen-like atoms II: spectral analysis of hydrogen and deuterium}

\author{David~M.~Jacobs}
\email{djacobs@norwich.edu}
\affiliation{Physics Department, Norwich University, 158 Harmon Dr, Northfield, VT 05663, USA}
\affiliation{CERCA, Physics Department,
Case Western Reserve University,
Cleveland, OH 44106, USA}

\date{\today}

\begin{abstract}
A long-distance effective theory of hydrogen-like atoms, dubbed the relativistic Ritz approach was recently introduced and some its theoretical consequences were explored. In this article, the relativistic Ritz approach is used to fit extant measurements of atomic hydrogen and deuterium transitions using information-theoretic analyses. As a result, the fine-structure constant ($\a$),  a fundamental parameter of the Standard Model, may be determined simultaneously with the ionization energies of hydrogen and deuterium, $E_I^{\text{(H)}}$ and $E_I^{\text{(D)}}$. The best hydrogen analysis yields $\a^{-1} = 137.035\,999\,185(25)$, in good agreement with the value obtained by other methods and without relying on a separately determined Rydberg constant. From the same analysis, I find that  $E_I^{\text{(H)}} = 13.598\,434\,599\,684(25)\,\text{eV}$,  an improvement of two orders of magnitude in precision compared to previous determinations and in agreement with the Standard Model prediction at 1.8 parts per trillion. The best deuterium analysis yields $E_I^{\text{(D)}} = 13.602\,134\,636\,543(31)\,\text{eV}$, agreeing with the Standard Model at 2.3 parts per trillion.  This study demonstrates how the relativistic Ritz approach can be used for testing the Standard Model with the spectra of hydrogen-like atoms. 
\end{abstract}
\maketitle

\section{Introduction}

The Standard Model of particle physics describes the fundamental interactions between particles on the smallest length scales. Despite its many successes, there are several apparent problems with the model. For example, it does not account for the observation of neutrino oscillations nor does it provide an explanation for the excess of matter over antimatter in the Universe \cite{Super-Kamiokande:1998kpq, Dine:2003ax}. Where gravity is concerned, there are other mysteries; for example, the mass in the Universe in the form of Standard Model particles was incapable of gravitationally coalescing on its own to form the astrophysical structure observed today \cite{Bertone:2016nfn}.
Testing the Standard Model at high precision thus remains a primary motivation to perform new physics experiments. High-energy particle colliders have, historically, provided a wealth of information about fundamental particle physics, but as the energies of these colliders has increased so has their cost, making the construction of new colliders harder to justify. 

In recent decades it has become apparent that high-precision experiments involving atoms and molecules are a complementary means to test the Standard Model and to probe for new physics \cite{safronova2018search}. Such experiments include measurements of the transitions between atomic states and are accompanied by either the emission or absorption of photons whose energy can be precisely measured. The most precise experiments are performed with  \emph{hydrogen-like} systems, which include deuterium and positronium\footnote{See, e.g.,  \cite{Cassidy:2018tgq} for recent developments on measurements of this system, composed of an electron bound with a positron.}.  Additionally, atomic measurements allow for the determination of some of the fundamental constants of the Standard Model.  In particular, the fine-structure constant 
\begin{equation}
\a=\frac{e^2}{4\pi\varepsilon_0 \hbar c}\,,
\end{equation}
quantifies the strength of the electromagnetic interaction and plays a pivotal role in physics and chemistry. Independent methods of measuring such constants are crucial both for building confidence in those values and also for testing the theory.

Bound-state quantum electrodynamics (BSQED) is the primary theoretical tool within the Standard Model for predicting the energy levels of atoms and for extracting fundamental constants from atomic spectra. Improving the precision of the theory is challenging (see, e.g., \cite{Indelicato_2019}), while spectroscopy may be subject to unanticipated sources of error \cite{Khabarova:2021trt}. Recently, a semi-empirical theory for hydrogen-like atoms  has been developed, dubbed the \emph{relativistic Ritz} approach \cite{Jacobs:2022mqm,Jacobs:2021xgp}. The approach rests on the fact that atomic effects beyond the lowest-order Coulomb interaction can be categorized by one of two types: (1) kinetic relativistic effects ($v/c>0$) and (2) shorter-ranged interactions, such as those associated with spin coupling, vacuum polarization, particle self-energy, and finite particle size. 

Within the Standard Model, the largest length scale associated with these effects is the (reduced) Compton wavelength of the electron, $r_\text{\tiny QED}=\hbar/(m_e c)$, whereas the orbital scale of the hydrogen atom in the $n$'th level is $r_\text{\tiny atom}=r_\text{\tiny QED}n^2/\a \simeq 137\, r_\text{\tiny QED}\,n^2$. By analyzing the quantum dynamics at long distance, i.e., $r_\text{\tiny atom}\gg r_\text{\tiny QED}$, it is possible to treat both the electron and nucleus as relativistic point-like objects and ``work inward", accounting for the omitted short-ranged interactions empirically through a low-energy series expansion of the quantum defect, $\de$. For hydrogen and deuterium, this long-distance criterion is true even for the ground state. The relativistic Ritz approach supersedes the older non-relativistic Ritz approach as a means to interpolate between and extrapolate beyond measured transitions\cite{Jacobs:2022mqm}; however, in this article it will be shown that the method may also be used to extract quantities relevant to the Standard Model, namely the atomic ionization energy, $E_I$, and the fine-structure constant, $\a$.

Given what will be demonstrated here, i.e., what the relativistic Ritz approach can accomplish with regard to spectroscopic data, it is useful to take stock of the methods that have been used hitherto to obtain ionization energies empirically, what the Standard Model predictions of those ionization energies are, as well as how some determinations of $\alpha$ are made.

To begin with, the canonical (nonrelativistic) Rydberg-Ritz fitting formula has historically been used to fit spectral data of alkali atoms, as well as atomic hydrogen and deuterium. This formula is
\begin{equation}\label{nonrel_Ryd_Ritz}
E_I - E_n = \f{hcR_\infty Z^2}{1+\f{m_e}{M_N}}\f{1}{\(n-\de\)^2}\,,
\end{equation}
where $E_n$ are the term energies, $hcR_\infty=m_e c^2 \a^2/2$ is the Rydberg-Hartree energy, $Z$ and $M_N$ are the nuclear charge and mass, respectively. Throughout this article, the defined CODATA \cite{Tiesinga:2021myr} values $h=6.626\,070\,15\times10^{-34} \text{\,J$\,\cdot\,$s}$, $c=299\,792\,458 \text{\,m/s}$, and $e=1.602\,176\,634\times10^{-19}\,\text{C}$ are used. The following (even-only) modified Ritz formula is often used as an ansatz for the quantum defect,
\begin{equation}
\de = c_0 + \f{c_2}{\(n-c_0\)^2} + \f{c_4}{\(n-c_0\)^4}+ \f{c_6}{\(n-c_0\)^6} + \dots
\end{equation}
In Ref. \cite{Kramida:2010lii}, this fitting formula was used to extract the atomic hydrogen ionization energy from spectral data contained within the ASD \cite{NIST_ASD},  
\begin{eqnarray}\label{E_I_old}
E_I^{\text{(H,data)}}&=&3\,288\, 086\, 856.8(7)\,\text{MHz} \times h\notag\\
&=&13.598\,434\,598(3)\,\text{eV}\,.
\end{eqnarray}
which is in agreement with the Standard Model prediction\cite{NIST_ASD},
\begin{equation}\label{QED_E0H}
E_I^{\text{(H,SM)}}=
13.598\,434\,599\,702(12)  \text{~eV}\,,
\end{equation}
 at the level of $2\times 10^{-10}$.  Likewise, the deuterium ionization energy is determined to be
\begin{eqnarray}\label{E_D_old}
E_I^{\text{(D,data)}}&=&3\,288\,981\,521.1(2.3)\,\text{MHz}\times h\notag\\
&=&13.602\,134\,633(10)\,\text{eV}\,.
\end{eqnarray}
which agrees with the Standard Model prediction,
\begin{equation}\label{QED_E0D}
E_I^{\text{(D,SM)}}=
13.602\,134\,636\,569(12)  \text{~eV}\,,
\end{equation}
 at the level of $7\times 10^{-10}$. As described in the sections that follow, I will improve upon the data-derived results of \eqref{E_I_old} and \eqref{E_D_old} by more than two orders of magnitude by using the relativistic generalization of equation \eqref{nonrel_Ryd_Ritz}. 
 
As for the fine-structure constant, $\a$,  in recent years, its most precise determinations have come by one of two means: (1) from the comparison of the experimentally determined electron g-factor \cite{Hanneke:2008tm} with the theoretical prediction \cite{Aoyama:2019ryr} and (2) atomic recoil experiments involving rubidium \cite{PhysRevLett.106.080801} or cesium \cite{Parker:2018vye}, combined with a chain of relative mass measurements \cite{huang2021ame} and the Rydberg constant \cite{Tiesinga:2021myr}, $R_\infty$. The Rydberg constant is determined from certain measured hydrogen and deuterium transitions, analyzed using BSQED. Recent determinations of the inverse fine-structure constant, $\a^{-1}$, are listed in Table \ref{table:alpha_inv_determinations}.

 \begin{table}[ht!]
\centering
\caption{A selection of fine-structure constant determinations.}
\begin{tabular}{c c c}  \hline\hline
$\a^{-1}$& Type & Reference\\ \hline
$137.035\,999\,166(15)$ & electron g-factor/QED & \cite{fan2022measurement, Aoyama:2019ryr}\\  
 $137.035\,999\,150(33)$ & electron g-factor/QED & \cite{Hanneke:2008tm, Aoyama:2019ryr}\\  
$137.035\,999\,046(27)$ & atom recoil/BSQED & \cite{Parker:2018vye}\\
$137.035\,999\,206(11)$ & atom recoil/BSQED & \cite{Morel:2020dww}\\ 
  \hline\hline
\end{tabular}
\label{table:alpha_inv_determinations}
\end{table}

Here I apply the relativistic Ritz approach to extant atomic hydrogen and deuterium spectroscopic data, demonstrating a new method for testing bound-state QED and the Standard Model, as well as checking for internal consistency of spectroscopic data. The former is done through the simultaneous determination of the ground state ionization energies and $\a$ with a precision that is competitive with the references mentioned above. In Section \ref{Prelims} I will summarize the approach and its input parameters for use in spectral analysis. In Sections \ref{Hydrogen_Analysis} and \ref{Deuterium_Analysis} I analyze  publicly available hydrogen and deuterium data, respectively. I conclude with a discussion in Section \ref{SummaryAndDiscussion}.

\sec{Preliminaries}\label{Prelims}

\ssec{Model summary, input parameters, and fitting procedure}
The energy levels of a hydrogen-like atom composed of two particles of mass $m_1$ and $m_2$ were shown in the relativistic Ritz approach \cite{Jacobs:2022mqm} to be
\begin{equation}\label{general_E_sol}
E_n=\sqrt{m_1^2+m_2^2+\f{2m_1m_2}{\sqrt{1+\(\f{Z\alpha}{n_\star}\)^2}}} - \Big(m_1+m_2\Big)\,,
\end{equation}
where the effective quantum number, 
\begin{equation}
n_\star=n-\de\,.
\end{equation}
The quantum defect, $\de$, causes an eigenfunction to deviate from its canonical ``pure Coulomb" form and accounts for omitted short-ranged interactions \cite{ritz1908new,hartree_1928,Seaton_1983,Jacobs:2019woc}. One may verify that, at lowest order in $Z\a/n_\star$, equation \eqref{general_E_sol} is equivalent to equation \eqref{nonrel_Ryd_Ritz}.  The standard defect ansatz is to posit that it admits a series expansion in energy, but the following \emph{modified} ansatz was shown to be equivalent and is significantly easier to use for data fitting
\begin{widetext}
\begin{multline}\label{defect_modified_ansatz}
\de=\de_{(0)\ell  j}+ \f{\de_{(2)\ell  j}}{\(n-\de_{(0)\ell  j}\)^2}+\f{\de_{(4)\ell  j}}{\(n-\de_{(0)\ell  j}\)^4} +\f{2\de_{(2)\ell  j}^2}{\(n-\de_{(0)\ell  j}\)^5}  +\f{\de_{(6)\ell  j}}{\(n-\de_{(0)\ell  j}\)^6}  +\f{6\de_{(2)\ell  j} \de_{(4)\ell  j}}{\(n-\de_{(0)\ell  j}\)^7}  +\f{\de_{(8)\ell  j}}{\(n-\de_{(0)\ell  j}\)^8} \\
+\f{4\de_{(4)\ell  j}^2 + 8\de_{(2)\ell  j} \de_{(6)\ell  j}}{\(n-\de_{(0)\ell  j}\)^9}
 +\f{\de_{(10)\ell  j}}{\(n-\de_{(0)\ell  j}\)^{10}} +\f{-40\de_{(2)\ell  j}^4 + 10\de_{(4)\ell  j} \de_{(6)\ell  j}+ 10\de_{(2)\ell  j} \de_{(8)\ell  j}}{\(n-\de_{(0)\ell  j}\)^{11}}+\dots\,.
\end{multline}
\end{widetext}
where $\ell$ and $j$ are the orbital and total electronic angular quantum numbers, respectively\footnote{Neither the total spin or total system angular momentum quantum number, $s$ or $f$, are here included because the hyperfine structure of most levels of hydrogen and deuterium has not been resolved.}. 

The input masses of the electron, proton, and deuteron depend on the atomic mass unit, $u$. At present, the best determination of $u$ comes from the rubidium recoil experiment of Ref \cite{Morel:2020dww}; relevant mass-related values are given in Table \ref{table:mass-related_parameters}. We consider excitation transitions, written in spectroscopic notation, $\[n \ell_j\]_\text{i}\to \[n \ell_j\]_\text{f}$; the frequencies ($\Delta\nu$) that correspond to these transitions are fit with the equation
\begin{equation}\label{transition_E_eqn}
h\Delta\nu = E_{n_\text{f} \ell_\text{f} j_\text{f}}-
\begin{cases}
E_0~~~~&(n_i=1)\\
E_{n_\text{i} \ell_\text{i} j_\text{i}}~~~~&(n_i\neq1)
\end{cases}
\end{equation}
where $E_{n\ell j}$ is the term energy defined in equation \eqref{general_E_sol}, $E_0$ is the ground state energy, the magnitude of which is the ground state ionization energy, $E_I$.
 
For each data fitting, we use the Levenburg-Marquardt algorithm, weighting each transition interval by the inverse-square of its measurement uncertainty. The statical analyses completed for this article were performed using the software \emph{Mathematica} and are included in a publicly available notebook \cite{jacobs_david_m_2022_7312051}.

\begin{table}[ht!]
\centering
\caption{Mass-related parameters}
\begin{tabular}{c c c}  \hline\hline
Quantity& Value & Reference$$\\ \hline
$h/m(^{87}\text{Rb})$ & $4.591\,359\,258\,90(65)\times10^{-9}\text{\,m$^2\,$s}^{-1}$ & \cite{Morel:2020dww}\\  
$m(^{87}\text{Rb})$ & $86.909\,180\,53\,10(60) \,u$ & \cite{huang2021ame}\\ 
$m(e)$ & $5.485\,799\,090\,65(16)\times10^{-4} \,u$ & \cite{sturm2014high}\\ 
$m(p)$ & $1.007\,276\,466\,583(32) \,u $ & \cite{heisse2017high}\\ 
$m(d)$ & $2.013\,553\,212\,535(17) \,u $ & \cite{rau2020penning}\\
  \hline\hline
\end{tabular}
\label{table:mass-related_parameters}
\end{table}

\ssec{Model Selection and Multi-model inference}\label{model_selec_inference_prelims}

The modified defect series, equation \eqref{defect_modified_ansatz}, must be truncated at some order for data analysis, therefore each choice of truncation represents a different model. In this article we only consider variations in truncation based on the angular momentum value, $\ell$; however, in the future, additional truncation choices should be considered, e.g., depending on $j$. At lowest order ($O_{\ell}=0$) only the defect parameter $\de_{(0)\ell j}$ is included, at next-to-lowest order ($O_{\ell}=1$) both $\de_{(0)\ell j}$ and $\de_{(2)\ell j}$ are included, and so on. There is no obvious place to truncate the series, and there may not be one, hence there are multiple (nested) models to consider, implying a \emph{model selection uncertainty}.  There are at least two distinct reasons for this. First, given ever-present experimental uncertainties, parsimony is a guiding principle in model selection which attempts to optimize for the inherent tradeoff between bias and parameter variance; too few parameters can result in significant bias and an overestimate of precision, while too many parameters can result in higher variance \cite{anderson2004model}. Finding the right balance requires care, as described in the paragraphs below.

The second reason for model selection uncertainty is because of the asymptotic foundation of the relativistic Ritz model. At the present time, it is not known if the series in equation \eqref{defect_modified_ansatz} converges, so it is for now conservatively assumed to be an asymptotic series. Such series are known to have an optimal point of truncation beyond which the approximation becomes less accurate; in fact, the series may diverge. However, useful information may be contained within \emph{all} terms of such a series \cite{bender1999advanced}, suggesting that analysis of models with more than the optimal number of parameters may be fruitful.

Ref \cite{anderson2004model} contains an excellent discussion of the use of the so-called second-order Akaike information criterion (AICc)  for comparing the goodness-of-fit of various models to a given data set. This information-theoretic criterion, a quantity that estimates the relative Kullback-Leibler entropy, is defined by
\begin{equation}
\text{AICc}_i = -2 \log{\cal L}(\hat{\theta}_i) + 2K_i\(\frac{N}{N-K_i-1}\)\,,
\end{equation}
where ${\cal L}(\hat{\theta}_i)$ is the maximum likelihood function, conditional on model $i$, for the set of best-fit model parameters, $\hat{\theta}_i$, $N$ is the sample size, and $K_i$ is the number of parameters in model $i$.  AICc, as opposed to Akaike's original AIC, is used whenever the data set is not sufficiently large; the rule of thumb is that AICc should be used whenever $N/K_i<40$ \cite{anderson2004model}.

 The ``best" model has the minimum value of AICc; however, the magnitudes of the AICc values are by themselves not meaningful. What matters is the AICc differences, defined by
\begin{equation}\label{AICc_diff_def}
\Delta_i\equiv \text{AICc}_i - \text{AICc}_{min}\,,
\end{equation}
and the \emph{evidence} of model $i$ depends exponentially on $\Delta_i$, namely it is proportional to $\exp{\(-\frac{1}{2}\Delta_i\)}$. However, it is often the case that many models fit to the same data will yield AICc values near to $\text{AICc}_{min}$, perhaps within 0 to 2 units, suggesting that the ``best" model may well vary -- sheerly by chance -- on a particular realization of data. In other words, if all data was re-collected and re-fit the AICc values would change and the best fit model may differ.  Wanting to draw robust conclusions from a single existing data set, therefore, leads us to average across multiple models. For this purpose, the \emph{Akaike weights} for model $i$ are introduced;
\begin{equation}
w_i = \frac{e^{-\frac{1}{2}\Delta_i}}{\sum_{j=1}^{R} e^{-\frac{1}{2}\Delta_j}}\,,
\end{equation}
where $R$ is the number of models under consideration. Note that by definition the Akaike weights sum to unity. The value of generic model parameter $\th_a$ will have a best estimate, conditional on model $i$, denoted here by $\hat{\theta}_{a,\,i}$. The model-averaged estimate for $\th_a$ is
\begin{equation}\label{weighted_average}
\langle\hat{\theta}_a\rangle = \sum_{i=1}^R \,w_i \,\hat{\theta}_{a,\,i}\,.
\end{equation}

Finally, we should seek an estimate of the error in \eqref{weighted_average} that is not conditional on a particular model, hence referred to as an \emph{unconditional} error estimate. The procedure described in \cite{anderson2004model}  includes both the conditional variance estimate, $\widehat{\text{var}}\(\hat{\theta}_{a,\,i}\)$, and a ``variance component" due to the fluctuation of the conditional parameter value around its model average, $\(\hat{\theta}_{a,\,i} - \langle\hat{\theta}_a\rangle\)^2$. The estimate of the unconditional standard error (se) is given by
\begin{equation}\label{unconditional_std_error}
\text{se}\(\hat{\theta}_a\) = \sum_{i=1}^R w_i\, \sqrt{\widehat{\text{var}}\(\hat{\theta}_{a,\,i}\) + \(\hat{\theta}_{a,\,i} - \langle\hat{\theta}_a\rangle\)^2}\,.
\end{equation}

Given the uncertainty inherent to model selection, as well as the psychological dangers associated with data analysis, e.g. data dredging, we proceed cautiously with several exploratory analyses involving subsets of hydrogen and deuterium transitions described in the sections below. 

\sec{Atomic hydrogen}\label{Hydrogen_Analysis}

The data used here are primarily found in the NIST Atomic Spectral Database (ASD) \cite{NIST_ASD}, which uses the 2010 hydrogen compilation \cite{Kramida:2010lii}. We make two updates, replacing the  $2S_{1/2}\to 8D_{5/2}$ transition with that from the more recent measurement of Ref. \cite{Brandt:2021yor} and adding the $1S_{1/2}\to 3S_{1/2}$ transition measurement from Ref. \cite{grinin2020two}. The ASD $2S_{1/2}\to 6S_{1/2}$ \& $2S_{1/2}\to 6D_{5/2}$ transitions are omitted from analysis, as they had not been directly observed\footnote{As explained in Ref. \cite{Kramida:2010lii}, those data reported in Ref. \cite{NIST_ASD} had only been determined from a combination of other measurements and from the (nonrelativistic) Ritz value of the $1S_{1/2}–3S_{1/2}$ frequency. I thank Alexander Kramida for clarifying this point. } at the time the 2010 compilation was published. As described in the line references in the ASD table, many lines identified as ``observed" are, in fact, only interpolations or extrapolations of measured transitions using the non-relativistic Ritz procedure, i.e. using \eqref{nonrel_Ryd_Ritz}; all such data are also omitted from analysis. The full list of transitions considered are shown in Tables \ref{table:H_transitions_part1} and \ref{table:H_transitions_part2}.

It is also important to emphasize what we fit here are the fine-structure transitions, i.e., without explicit consideration of interactions that involve the nuclear spin. However, many measurements require theoretical corrections to obtain the corresponding fine-structure transition. In this sense, the data used here generally represent a combination of experimental results with statistical mechanical and bound-state QED theory corrections. 

\subsection{Exploratory hydrogen analysis: $S_{1/2}$ \& $D_{5/2}$ states}\label{Exploratory_Hyd_SD_Analysis}

Here we consider only the transitions between the $S_{1/2}$ and $D_{5/2}$ states, of which there are only 10 measurements, summarized below in Table \ref{tab:std_res_expl_SD}. The relativistic Ritz model at various points of truncation of the modified defect (equation \eqref{defect_modified_ansatz}) are labeled  RR$O_SO_D$.  A selection of models were fit to this data and the AICc differences, $\Delta_i$, are reported in Table \ref{table:Exploratory_hydrogen_AICc}. Given the exponential dependence of the relative evidence on $\Delta_i$, model RR10  --  a next-to-lowest order fit for $S$ states and a lowest order fit for $D$ states -- is by far the best. The parameter fits from the RR10 model analysis are reported in Table  \ref{table:Exploratory_hydrogen_RR10}.   Note that there are significant correlations amongst the model parameters, which are quantified in the correlation matrix shown in Table \ref{table:Exploratory_hydrogen_RR10_Correlation_Matrix}.

  \begin{table}[htp]
  \centering
   \caption{List of hydrogen transition data used for exploratory fitting. Alternate references are given for data taken from other than the compilation in Ref.  \cite{NIST_ASD}.}
\begin{tabular}{c |  c c c}
Number& Transition &  Frequency (GHz)  \\\hline
1& $ 1 \text{S}_{\text{1/2}}\to 3 \text{S}_{\text{1/2}} $ & $ 2\,922\,743.278\,665\,79(72)$& \cite{grinin2020two} \\
 2& $ 1 \text{S}_{\text{1/2}}\to 2 \text{S}_{\text{1/2}} $ & $ 2\,466\,061.413\,187\,070(30)$ \\
 3& $ 2 \text{S}_{\text{1/2}}\to 12 \text{D}_{\text{5/2}} $ & $ 799\,191.727\,4038(64)$ \\
 4& $2 \text{S}_{\text{1/2}}\to 10 \text{D}_{\text{5/2}} $ & $ 789\,144.886\,412(39) $\\
 5& $ 2 \text{S}_{\text{1/2}}\to 8 \text{D}_{\text{5/2}} $ & $ 770\,649.561\,5709(20) $&\cite{Brandt:2021yor}  \\
 6& $ 2 \text{S}_{\text{1/2}}\to 8 \text{S}_{\text{1/2}} $ & $ 770\,649.350\,0121(99)$ \\
7& $ 2 \text{S}_{\text{1/2}}\to 4 \text{D}_{\text{5/2}} $ & $ 616\,521.843\,441(24) $ \\
  8& $ 2 \text{S}_{\text{1/2}}\to 4 \text{S}_{\text{1/2}} $ & $ 616\,520.150\,637(10) $ \\
   9& $ 3 \text{S}_{\text{1/2}}\to 3 \text{D}_{\text{5/2}} $ & $ 4.013\,162(48) $ \\
    10& $ 4 \text{S}_{\text{1/2}}\to 4 \text{D}_{\text{5/2}} $ & $ 1.692\,98(38) $ \\
  \hline\hline
\end{tabular}
\label{tab:std_res_expl_SD}
\end{table}

\begin{table}[H]
\centering
\caption{Second-order Akaike information criterion differences, $\Delta_i$, for the exploratory analysis of $S_{1/2}\to S_{1/2} ~\&~ D_{5/2}$ transitions.}
\begin{tabular}{l  | c c c c}  \hline\hline
&&$O_{D}$\\
$O_{S}$ &0 (LO)&1 (NLO) &2 (NNLO) &3 (N$^3$LO) \\ \hline
$0$ (LO) & 35.5 & 47.2 & 78.3 & 162. \\
$1$ (NLO) &  0 & 27.5 & 106. & $\infty$  \\
$2$ (NNLO) &  15.8 & 90.4 & $\infty$  & $\infty$  \\
$3$ (N$^3$LO) &  108. & $\infty$  & $\infty$  & --  \\ \hline\hline
\end{tabular}
\label{table:Exploratory_hydrogen_AICc}
\end{table}

\begin{table}[H]
\centering
\caption{Fit parameters for model RR10 used in the exploratory analysis.}
\begin{tabular}{c | c}  \hline\hline
$\de_{(0) 0\f{1}{2}}$ & $2.5344(1)\times10^{-5}$\\  
$\de_{(2) 0\f{1}{2}}$ & $4.8(2)\times10^{-8}$\\  
$\de_{(0) 2\f{5}{2}}$ & $8.8724(9)\times10^{-6}$\\   \hline
$E_0^\text{(H)}\,\[\text{eV}\]$ & $-13.598\,434\,599\,543(73)$\\  
$\a^{-1}$  & $137.035\,999\,298(38)$\\  \hline\hline
\end{tabular}
\label{table:Exploratory_hydrogen_RR10}
\end{table}

\begin{table}[H]
\centering
\caption{Correlation Matrix for the RR10 model in the exploratory analysis.}
\begin{tabular}{c | c c c c c}  \hline\hline
&$\de_{(0) 0\f{1}{2}}$ & $\de_{(2) 0\f{1}{2}}$& $\de_{(0) 2\f{5}{2}}$& $E_0$& $\a^{-1}$\\  \hline
$\de_{(0) 0\f{1}{2}}$ &1 & &  & &\\  
$\de_{(2) 0\f{1}{2}}$ &-0.996  &1 & & &\\  
$\de_{(0) 2\f{5}{2}}$ &0.9612&-0.9657 &1  & &\\  
$E_0$ &0.9558 &-0.9474 &0.8645 &1  &\\  
$\a^{-1}$  &0.9995 &-0.9980 &0.9550 &0.9646 &1 \\  \hline\hline
\end{tabular}
\label{table:Exploratory_hydrogen_RR10_Correlation_Matrix}
\end{table}

 As a check of the conditional sampling error estimates in Table \ref{table:Exploratory_hydrogen_RR10}, we perform a bootstrap analysis, resampling and replacing each of the 10 transition measurements with a random value drawn from a normal distribution whose mean is the measured value and standard deviation is equal to the measurement uncertainty. Ten thousand ($B=10,000$) such re-samplings were performed and analyzed with the RR10 model, the results of which give standard deviations $\sigma_{E_0} = 4.3\times 10^{-10}\,\text{eV}$ and $\sigma_{\a^{-1}} =2.2\times10^{-8}$, in reasonable agreement with the conditional sampling error estimates shown in Table \ref{table:Exploratory_hydrogen_RR10}.

In addition to these statistical considerations, the mass-related inputs (Table \ref{table:mass-related_parameters}) have uncertainties that must be propagated. To account for this, we perform a Monte Carlo analysis, resampling and replacing each of the 4 mass-related measurements with a random value drawn from a normal distribution whose mean and standard deviation are given by its measured value and uncertainty, respectively. Ten thousand ($B=10,000$) such re-samplings were performed and analyzed with the RR10 model, giving standard deviations $\sigma_{E_0} \leq10^{-16}\,\text{eV}$ (negligible) and $\sigma_{\a^{-1}} =1.1\times10^{-8}$. The latter value is, not coincidentally, equal to the uncertainty in $\sigma_{\a^{-1}}$ claimed by the authors of Ref. \cite{Morel:2020dww}.


Finally, from this exploratory analysis
\begin{eqnarray}\label{exploratory_H_alpha_inv_final}
\a^{-1} &=&  137.035\,999\,298 (38)_\text{data} (11)_\text{mass}\notag\\
&=&  137.035\,999\,298 (40)
\end{eqnarray}
where the subscript ``data" denotes the statistical uncertainty due to transition data fitting and model averaging and ``mass" denotes uncertainty due to uncertainty in the mass-related inputs. The ionization energy of hydrogen from this analysis is
\begin{equation}\label{exploratory_E_I_H_final}
E_I^{\text{(H)}} = 13.598\,434\,599\,543(73)\,\text{eV}\,.
\end{equation}

There is a discrepancy of approximately $\simeq2.5\s$ between \eqref{exploratory_H_alpha_inv_final} and \eqref{exploratory_E_I_H_final} and their expected values in Table \ref{table:alpha_inv_determinations} and equation \eqref{QED_E0H}, respectively. Notably, our determination of $\a^{-1}$ here is systematically too high, but this is expected. When fewer parameters are used $\a^{-1}$ will be biased toward larger values, as found in the theoretical analysis in Ref \cite{Jacobs:2022mqm}. Through the inclusion of more transitions, it becomes feasible that models with more parameters are favored, thus leading to a more reliable determination of $\a$ and $E_I$. We address this in the following subsections.

\ssec{Initial hydrogen $S,P,$ \& $D$-channel analysis}\label{Uncut_H_SPD_analysis}

Here we analyze the 60 measured hydrogen transitions coming primarily the NIST ASD 2010 hydrogen compilation \cite{NIST_ASD}, which includes all transitions between the 5 angular momentum channels $S_{1/2},P_{1/2},P_{3/2},D_{3/2}$, and $D_{5/2}$. These data are summarized in Tables \ref{table:H_transitions_part1} and \ref{table:H_transitions_part2}. A selection of AICc differences are given in Tables \ref{table:H_SPD_uncut_OD=0}, \ref{table:H_SPD_uncut_OD=1}, and \ref{table:H_SPD_uncut_OD=2}.

\begin{table}[p!]

\centering
\caption{Hydrogen transitions. References are given for those taken from sources other than the compilation in Ref.  \cite{NIST_ASD}.}
\begin{tabular}{c  |  c c c}
Number& Transition &  Frequency (GHz)  \\\hline
 1 & $ 1 \text{S}_{\text{1/2}}\to 12 \text{P}_{\text{3/2}} $ & $ 3\,265\,251(71) $ \\
 2 & $ 1 \text{S}_{\text{1/2}}\to 12 \text{P}_{\text{1/2}} $ & $ 3\,265\,251(71) $\\
 3 & $ 1 \text{S}_{\text{1/2}}\to 11 \text{P}_{\text{3/2}} $ & $ 3\,260\,921(71) $ \\
 4 & $ 1 \text{S}_{\text{1/2}}\to 11 \text{P}_{\text{1/2}} $ & $ 3\,260\,921(71) $ \\
 5 & $ 1 \text{S}_{\text{1/2}}\to 10 \text{P}_{\text{3/2}} $ & $ 3\,255\,182(71) $ \\
 6 & $ 1 \text{S}_{\text{1/2}}\to 10 \text{P}_{\text{1/2}} $ & $ 3\,255\,182(71) $ \\
 7 & $ 1 \text{S}_{\text{1/2}}\to 9 \text{P}_{\text{3/2}} $ & $ 3\,247\,491(70) $ & \\
 8 & $ 1 \text{S}_{\text{1/2}}\to 9 \text{P}_{\text{1/2}} $ & $ 3\,247\,491(70) $ & \\
 9 & $ 1 \text{S}_{\text{1/2}}\to 8 \text{P}_{\text{3/2}} $ & $ 3\,236\,699(70) $ & \\
 10 & $ 1 \text{S}_{\text{1/2}}\to 8 \text{P}_{\text{1/2}} $ & $ 3\,236\,699(70) $ \\
 11 & $ 1 \text{S}_{\text{1/2}}\to 7 \text{P}_{\text{3/2}} $ & $ 3\,220\,981(69) $ \\
 12 & $ 1 \text{S}_{\text{1/2}}\to 7 \text{P}_{\text{1/2}} $ & $ 3\,220\,981(69) $ \\
 13 & $ 1 \text{S}_{\text{1/2}}\to 6 \text{P}_{\text{3/2}} $ & $ 3\,196\,760(72) $ \\
 14 & $ 1 \text{S}_{\text{1/2}}\to 6 \text{P}_{\text{1/2}} $ & $ 3\,196\,760(72) $ \\
 15 & $ 1 \text{S}_{\text{1/2}}\to 5 \text{P}_{\text{3/2}} $ & $ 3\,156\,567(13) $ \\
 16 & $ 1 \text{S}_{\text{1/2}}\to 5 \text{P}_{\text{1/2}} $ & $ 3\,156\,567(13) $ \\
 17 & $ 1 \text{S}_{\text{1/2}}\to 4 \text{P}_{\text{3/2}} $ & $ 3\,082\,569(60) $ \\
 18 & $ 1 \text{S}_{\text{1/2}}\to 4 \text{P}_{\text{1/2}} $ & $ 3\,082\,569(60) $ \\
 19 & $ 1 \text{S}_{\text{1/2}}\to 3 \text{S}_{\text{1/2}} $ & $ 2\,922\,743.278\,665\,79(72)$& \cite{grinin2020two} \\
 20 & $ 1 \text{S}_{\text{1/2}}\to 3 \text{P}_{\text{3/2}} $ & $ 2\,922\,728.6(8.5) $ \\
 21 & $ 1 \text{S}_{\text{1/2}}\to 3 \text{P}_{\text{1/2}} $ & $ 2\,922\,728.6(8.5) $ \\
 22 & $ 1 \text{S}_{\text{1/2}}\to 2 \text{P}_{\text{3/2}} $ & $ 2\,466\,068.0(4.1) $ \\
 23 & $ 1 \text{S}_{\text{1/2}}\to 2 \text{P}_{\text{1/2}} $ & $ 2\,466\,068.0(4.1) $ \\
 24 & $ 1 \text{S}_{\text{1/2}}\to 2 \text{S}_{\text{1/2}} $ & $ 2\,466\,061.413\,187\,070(30)$ \\
 25 & $ 2 \text{S}_{\text{1/2}}\to 12 \text{D}_{\text{5/2}} $ & $ 799\,191.727\,4038(64)$ \\
 26 & $ 2 \text{S}_{\text{1/2}}\to 12 \text{D}_{\text{3/2}} $ & $ 799\,191.710\,473(11) $ \\
 27 & $ 2 \text{S}_{\text{1/2}}\to 10 \text{D}_{\text{5/2}} $ & $ 789\,144.886\,412(39) $\\
 28 & $ 2 \text{S}_{\text{1/2}}\to 8 \text{D}_{\text{5/2}} $ & $ 770\,649.561\,5709(20) $&\cite{Brandt:2021yor}  \\
 29 & $ 2 \text{S}_{\text{1/2}}\to 8 \text{D}_{\text{3/2}} $ & $ 770\,649.504\,4499(79)$  \\
 30 & $ 2 \text{S}_{\text{1/2}}\to 8 \text{S}_{\text{1/2}} $ & $ 770\,649.350\,0121(99)$ \\
 \hline \hline
\end{tabular}
\label{table:H_transitions_part1}
\end{table}

\begin{table}[p!]
\centering
\caption{Continuation of Table \ref{table:H_transitions_part1}.}
\begin{tabular}{c  |  c c c}
Number& Transition &  Frequency (GHz)  \\\hline
  31 & $ 2 \text{S}_{\text{1/2}}\to 4 \text{D}_{\text{5/2}} $ & $ 616\,521.843\,441(24) $ \\
 32 & $ 2 \text{S}_{\text{1/2}}\to 4 \text{P}_{\text{3/2}} $ & $ 616\,521.388\,672(10) $ \\
 33 & $ 2 \text{S}_{\text{1/2}}\to 4 \text{S}_{\text{1/2}} $ & $ 616\,520.150\,637(10) $ \\
 34 & $ 2 \text{S}_{\text{1/2}}\to 4 \text{P}_{\text{1/2}} $ & $ 616\,520.017\,568(15) $ \\
 35 & $ 2 \text{P}_{\text{1/2}}\to 3 \text{D}_{\text{3/2}} $ & $ 456\,685.8528(17) $ \\
 36 & $ 2 \text{S}_{\text{1/2}}\to 3 \text{P}_{\text{3/2}} $ & $ 456\,684.800\,11(28) $ \\
 37 & $ 2 \text{S}_{\text{1/2}}\to 3 \text{P}_{\text{1/2}} $ & $ 456\,681.549\,89(28) $ \\
 38 & $ 2 \text{P}_{\text{3/2}}\to 3 \text{D}_{\text{5/2}} $ & $ 456\,675.9681(42) $ \\
 39 & $ 2 \text{P}_{\text{1/2}}\to 2 \text{P}_{\text{3/2}} $ & $ 10.969\,13(10) $ \\
 40 & $ 2 \text{S}_{\text{1/2}}\to 2 \text{P}_{\text{3/2}} $ & $ 9.911\,202(13) $ \\
 41 & $ 3 \text{S}_{\text{1/2}}\to 3 \text{D}_{\text{5/2}} $ & $ 4.013\,162(48) $ \\
 42 & $ 3 \text{P}_{\text{1/2}}\to 3 \text{P}_{\text{3/2}} $ & $ 3.250\,31(39) $ \\
 43 & $ 3 \text{P}_{\text{1/2}}\to 3 \text{D}_{\text{3/2}} $ & $ 3.2449(32) $ \\
 44 & $ 3 \text{S}_{\text{1/2}}\to 3 \text{P}_{\text{3/2}} $ & $ 2.9334(11) $ \\
 45 & $ 3 \text{S}_{\text{1/2}}\to 3 \text{D}_{\text{3/2}} $ & $ 2.929\,95(86) $ \\
 46 & $ 4 \text{S}_{\text{1/2}}\to 4 \text{D}_{\text{5/2}} $ & $ 1.692\,98(38) $ \\
 47 & $ 4 \text{P}_{\text{1/2}}\to 4 \text{D}_{\text{3/2}} $ & $ 1.3711(12) $ \\
 48 & $ 4 \text{P}_{\text{1/2}}\to 4 \text{P}_{\text{3/2}} $ & $ 1.370\,86(25) $ \\
 49 & $ 4 \text{S}_{\text{1/2}}\to 4 \text{P}_{\text{3/2}} $ & $ 1.237\,79(31) $ \\
 50 & $ 4 \text{S}_{\text{1/2}}\to 4 \text{D}_{\text{3/2}} $ & $ 1.2352(20) $ \\
 51 & $ 3 \text{D}_{\text{3/2}}\to 3 \text{D}_{\text{5/2}} $ & $ 1.0831(31) $ \\
 52 & $ 3 \text{P}_{\text{3/2}}\to 3 \text{D}_{\text{5/2}} $ & $ 1.0780(12) $ \\
 53 & $ 2 \text{P}_{\text{1/2}}\to 2 \text{S}_{\text{1/2}} $ & $ 1.057\,8470(9.0) $ \\
 54 & $ 5 \text{P}_{\text{1/2}}\to 5 \text{D}_{\text{3/2}} $ & $ 0.7037(66) $ \\
 55 & $ 5 \text{S}_{\text{1/2}}\to 5 \text{P}_{\text{3/2}} $ & $ 0.622(12) $ \\
 56 & $ 4 \text{P}_{\text{3/2}}\to 4 \text{D}_{\text{5/2}} $ & $ 0.4557(16) $ \\
 57 & $ 3 \text{P}_{\text{1/2}}\to 3 \text{S}_{\text{1/2}} $ & $ 0.314\,819(50) $ \\
 58 & $ 5 \text{P}_{\text{3/2}}\to 5 \text{D}_{\text{5/2}} $ & $ 0.2320(31) $ \\
 59 & $ 4 \text{P}_{\text{1/2}}\to 4 \text{S}_{\text{1/2}} $ & $ 0.133\,18(59) $ \\
 60 & $ 3 \text{D}_{\text{3/2}}\to 3 \text{P}_{\text{3/2}} $ & $ 0.005\,45(89) $ \\ \hline \hline
\end{tabular}
\label{table:H_transitions_part2}
\end{table}

\begin{table}[H]
\centering
\caption{Second-order Akaike information criterion differences, $\Delta_i$, for the initial analysis of $S,P,$ and $D$-channel transitions in hydrogen. Here $O_D=0$. Bolded values identify the models used for averaging.}
\begin{tabular}{c  | c c c c c }  \hline\hline
&&&$O_{P}$\\
$O_{S}$ & 0& 1 & 2 & 3 & 4\\ \hline
0 & 368.8 & 257.2 & 261.7 & 268.9 & 276.8 \\
1 &  30.1 & 13.1 & 17.5 & 24.8 & 33.1 \\
2 &  32.9 & {\bf 0.} & {\bf 3.6} & 11.2 & 19.9 \\
3 &  35.6 & {\bf 3.5} & {\bf 7.5} & 15.5 & 24.6 \\
4 &  38.1 & 5.5 & 9.7 & 18.1 & 27.7 \\ \hline\hline
\end{tabular}
\label{table:H_SPD_uncut_OD=0}
\end{table}

\begin{table}[H]
\centering
\caption{Second-order Akaike information criterion differences, $\Delta_i$, for the initial analysis of $S,P,$ and $D$-channel transitions in hydrogen. Here $O_D=1$. Bolded values identify the models used for averaging.}
\begin{tabular}{c  | c c c c c }  \hline\hline
&&&$O_{P}$\\
$O_{S}$ & 0& 1 & 2 & 3 & 4\\ \hline
0 & 374.3 & 236.6 & 239.6 & 247.5 & 256.3 \\
1 &  19.2 & 14.0 & 18.7 & 26.8 & 35.9 \\
2 &  22.4 & {\bf 5.3} & {\bf 7.0} & 15.4 & 25.0 \\
3 &  25.5 & {\bf 9.0} & {\bf 10.6} & 19.5 & 29.6 \\
4 &  28.0 & 11.3 & 13.1 & 22.4 & 33.1 \\ \hline\hline
\end{tabular}
\label{table:H_SPD_uncut_OD=1}
\end{table}

\begin{table}[H]
\centering
\caption{Second-order Akaike information criterion differences, $\Delta_i$, for the initial analysis of $S,P,$ and $D$-channel transitions in hydrogen. Here $O_D=2$.}
\begin{tabular}{c  | c c c c c }  \hline\hline
&&&$O_{P}$\\
$O_{S}$ & 0& 1 & 2 & 3 & 4\\ \hline
0 & 379.4 & 240.9 & 244.2 & 252.9 & 262.6 \\
1 &  14.6 & 8.2 & 12.8 & 21.7 & 31.8 \\
2 &  16.3 & 8.1 & 11.5 & 21.0 & 31.7 \\
3 &  13.7 & 10.9 & 16.1 & 26.1 & 37.4 \\
4 &  16.1 & 13.4 & 19.1 & 29.5 & 41.4 \\ \hline\hline
\end{tabular}
\label{table:H_SPD_uncut_OD=2}
\end{table}

Despite the well-established procedures for model averaging and error estimation described in Section \ref{model_selec_inference_prelims}, there is still freedom in the decision of which models to include in this analysis.  At the very least, it would seem that model RR210 should be included in the average, but what others? Models RR220 and RR310 have only moderately large AICc differences, so they should also be included. Considering that this would represent a kind of exploration in the number of $S$- and $P$-channel defect parameters, it seems prudent to also consider model RR320, as well as a variation in the number of $D$-state parameters. For this reason, we also include models RR211, RR221, RR311, and RR321. \emph{This creates a contiguous and compact exploration of the $(O_S,O_P, O_D)$ parameter space}. These 8 models are identified with bold text in Tables \ref{table:H_SPD_uncut_OD=0} and \ref{table:H_SPD_uncut_OD=1}.  The Akaike weights for the set of these 8 models, along with a selection of parameter values are summarized in Table \ref{table:A-weight-FS-EIH_initial_HSPD}.

\begin{centering}

\begin{table}[H]
\centering
\caption{Akaike weights and some parameter fits for the initial (uncut) hydrogen $S,P,$ and $D$-channel analysis.}
\begin{tabular}{c  |  c c c}  \hline\hline
Model & $w_i$ & $\a^{-1}$ & $E_I^{\text{(H)}} \text{eV}$ \\ \hline
RR210 & $0.6774$ &  $137.035\,999\,248(21)$ & $13.598\,434\,599\,625(38)$  \\  
RR220 & $0.1100$ &  $137.035\,999\,246(22)$ & $13.598\,434\,599\,630(39)$  \\  
RR310 & $0.1178$ &  $137.035\,999\,248(22)$ & $13.598\,434\,599\,626(40)$  \\  
RR320 & $0.0161$ &  $137.035\,999\,245(22)$ & $13.598\,434\,599\,631(40)$  \\  
RR211 & $0.0475$ &  $137.035\,999\,207(40)$ & $13.598\,434\,599\,659(48)$  \\  
RR221 & $0.0203$ &  $137.035\,999\,170(46)$ & $13.598\,434\,599\,697(53)$  \\  
RR311 & $0.0075$ &  $137.035\,999\,203(42)$ & $13.598\,434\,599\,660(48)$  \\  
RR321 & $0.0033$ &  $137.035\,999\,160(49)$ & $13.598\,434\,599\,702(54)$  \\  \hline\hline
\end{tabular}
\label{table:A-weight-FS-EIH_initial_HSPD}
\end{table}

\end{centering}

Going through the model averaging and uncertainty analysis described in Section \ref{model_selec_inference_prelims}, we determine
\begin{eqnarray}\label{alpha_inv_initial_H_SPD}
\a^{-1} &=&  137.035\,999\,244 (25)_\text{data} (11)_\text{mass}\notag\\
&=&  137.035\,999\,244 (27)\,,
\end{eqnarray}
which is approximately $1.5\sigma$ discrepant from the consensus of values in \ref{table:alpha_inv_determinations}. The hydrogen ground state ionization energy is simultaneously determined to be
\begin{equation}\label{EHI_initial_H_SPD}
E_I^{\text{(H)}} = 13.598\,434\,599\,629(41)\,\text{eV}\,,
\end{equation}
which is approximately $1.5\sigma$ discrepant from the Standard Model prediction in equation \eqref{QED_E0H}.

It is at this point worth evaluating some measure of goodness-of-fit. This is done by considering the fit residuals, defining the residual of the $i$'th transition as
\begin{equation}
\varepsilon_i = \Delta \nu_{\text{measured}, i} - \Delta \nu_{\text{model}, i}\,.
\end{equation}
One approach would be to consider these residuals normalized to the measurement uncertainty of each transition. Doing so results in a mean of $-0.126$ and a standard deviation of $0.787$. Another option is to consider the so-called standardized residuals, defined to have a mean of $0$ and standard deviation equal to unity. The advantage of the standardized residuals is that they take into account the leverage of each data point, making data that deviate from the model prediction stand out when those data have a larger influence on the fit. The standardized residuals from the RR210 fit are shown in Figure \ref{Hyd_uncut_SPD_Std_Residuals_1-60}. For every model, the transition $2S_{1/2}\to3P_{1/2}$ stands out as significantly discrepant; in all models the standardized residual for this datum is lower than -3.0. Although such discrepancies are expected to occur due to random measurement error, a discrepancy at or exceeding $3\s$ is expected with an occurrence of 0.3\%, or an average of only 0.18 out of 60. The anomalous nature of this data point warrants another analysis with it removed, which we do in the following section.

\emph{All workers following this approach are advised to be 
both cautious when pursuing such investigations and transparent about their methodologies when publicizing their results.}

\begin{centering}
\begin{widetext}
~
\begin{figure}[H]
  \begin{center}
    \includegraphics[scale=0.8]{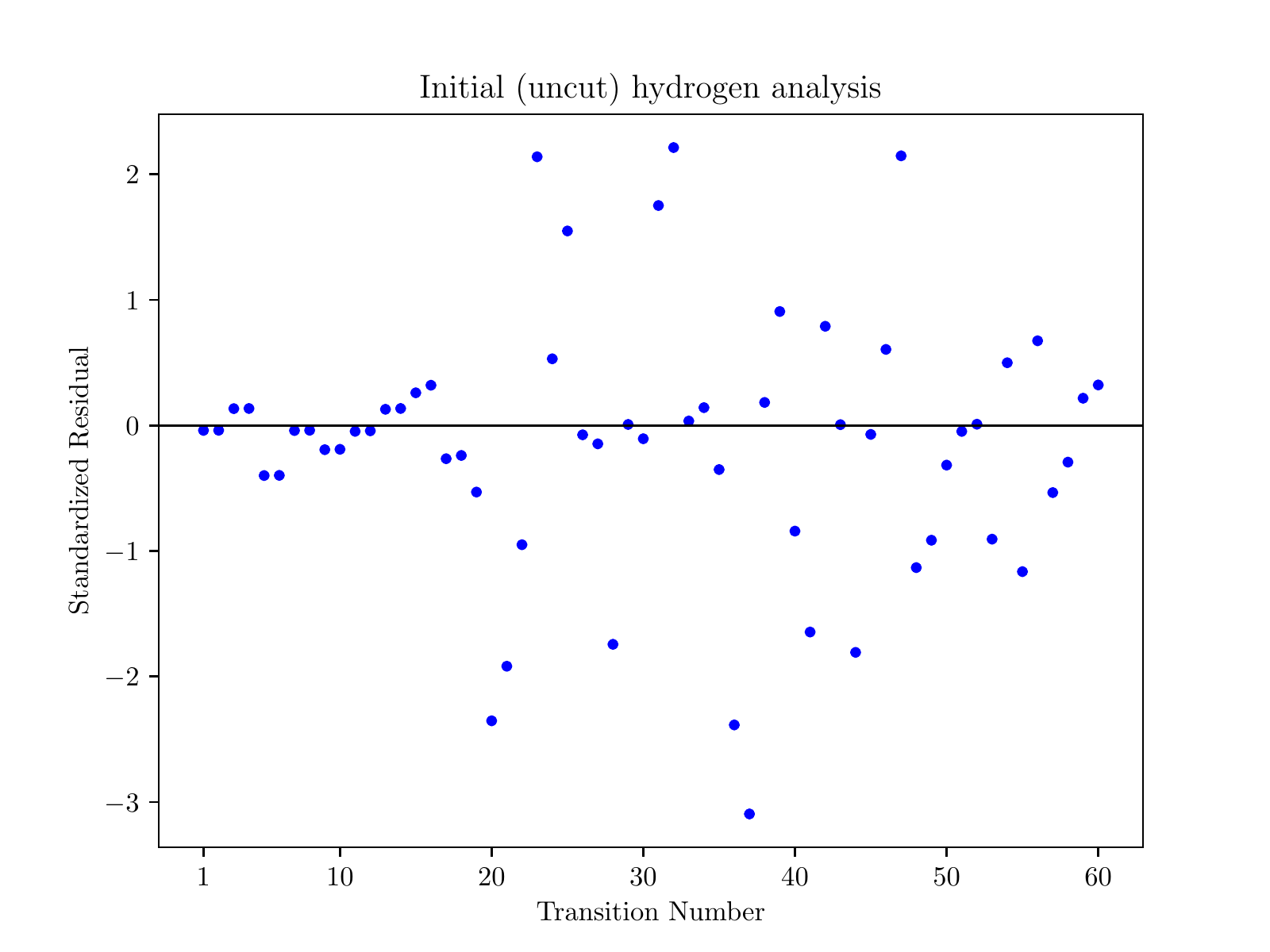}
  \end{center}
  \caption{Standardized residuals from the initial (uncut) hydrogen analysis. Transitions identified in Tables \ref{table:H_transitions_part1} and \ref{table:H_transitions_part2}.}
  \label{Hyd_uncut_SPD_Std_Residuals_1-60}
  \end{figure}

\end{widetext}
\end{centering}

\ssec{Hydrogen analysis without $2S_{1/2}\to3P_{1/2}$}\label{firstcut_H_SPD_analysis_no_2S3P}

Here we repeat the analysis of Section \ref{Uncut_H_SPD_analysis} without the $2S_{1/2}\to3P_{1/2}$ transition.  The 8 best models remain the same as in the previous section. The Akaike weights and some parameter values are given in Table \ref{table:A-weight-FS-EIH_first cut_HSPD}.

\begin{centering}
~
\begin{table}[H]
\centering
\caption{Akaike weights and some parameter fits for the hydrogen $S,P,$ and $D$-channel analysis, removing the $2S_{1/2}\to3P_{1/2}$ transition. }
\begin{tabular}{c  |  c c c}  \hline\hline
Model & $w_i$ & $\a^{-1}$ & $E_I^{\text{(H)}} \text{eV}$ \\ \hline
RR210 & $0.6041$ &  $137.035\,999\,246(19)$ & $13.598\,434\,599\,629(35)$  \\  
RR220 & $0.1306$ &  $137.035\,999\,246(19)$ & $13.598\,434\,599\,630(35)$  \\  
RR310 & $0.1036$ &  $137.035\,999\,246(20)$ & $13.598\,434\,599\,630(36)$  \\  
RR320 & $0.0187$ &  $137.035\,999\,245(20)$ & $13.598\,434\,599\,631(36)$  \\  
RR211 & $0.0821$ &  $137.035\,999\,197(36)$ & $13.598\,434\,599\,670(43)$  \\  
RR221 & $0.0402$ &  $137.035\,999\,170(41)$ & $13.598\,434\,599\,697(47)$  \\  
RR311 & $0.0138$ &  $137.035\,999\,192(37)$ & $13.598\,434\,599\,671(43)$  \\  
RR321 & $0.0070$ &  $137.035\,999\,160(43)$ & $13.598\,434\,599\,702(48)$  \\  \hline\hline
\end{tabular}
\label{table:A-weight-FS-EIH_first cut_HSPD}
\end{table}
\end{centering}

 Performing the averaging and uncertainty analysis again, we determine
\begin{eqnarray}\label{alpha_inv_2nd_H_SPD}
\a^{-1} &=&  137.035\,999\,238 (27)_\text{data} (11)_\text{mass}\notag\\
&=&  137.035\,999\,244 (29)\,,
\end{eqnarray}
and
\begin{equation}\label{EHI_2nd_H_SPD}
E_I^{\text{(H)}} = 13.598\,434\,599\,636(40)\,\text{eV}\,,
\end{equation}
the latter of which is a marginal improvement over the results from the previous section. However, the standardized residuals, which are shown in Figure \ref{Hyd_firstcut_SPD_Std_Residuals_1-59}, show that both the $2S_{1/2}\to3P_{3/2}$ and $1S_{1/2}\to3P_{3/2}$ transitions are discrepant at about $2.6\sigma$. Again, this is statistically unlikely to occur with only 59 data points and motivates their removal in another re-analysis.

\begin{centering}
\begin{widetext}

\begin{figure}[H]
  \begin{center}
    \includegraphics[scale=.8]{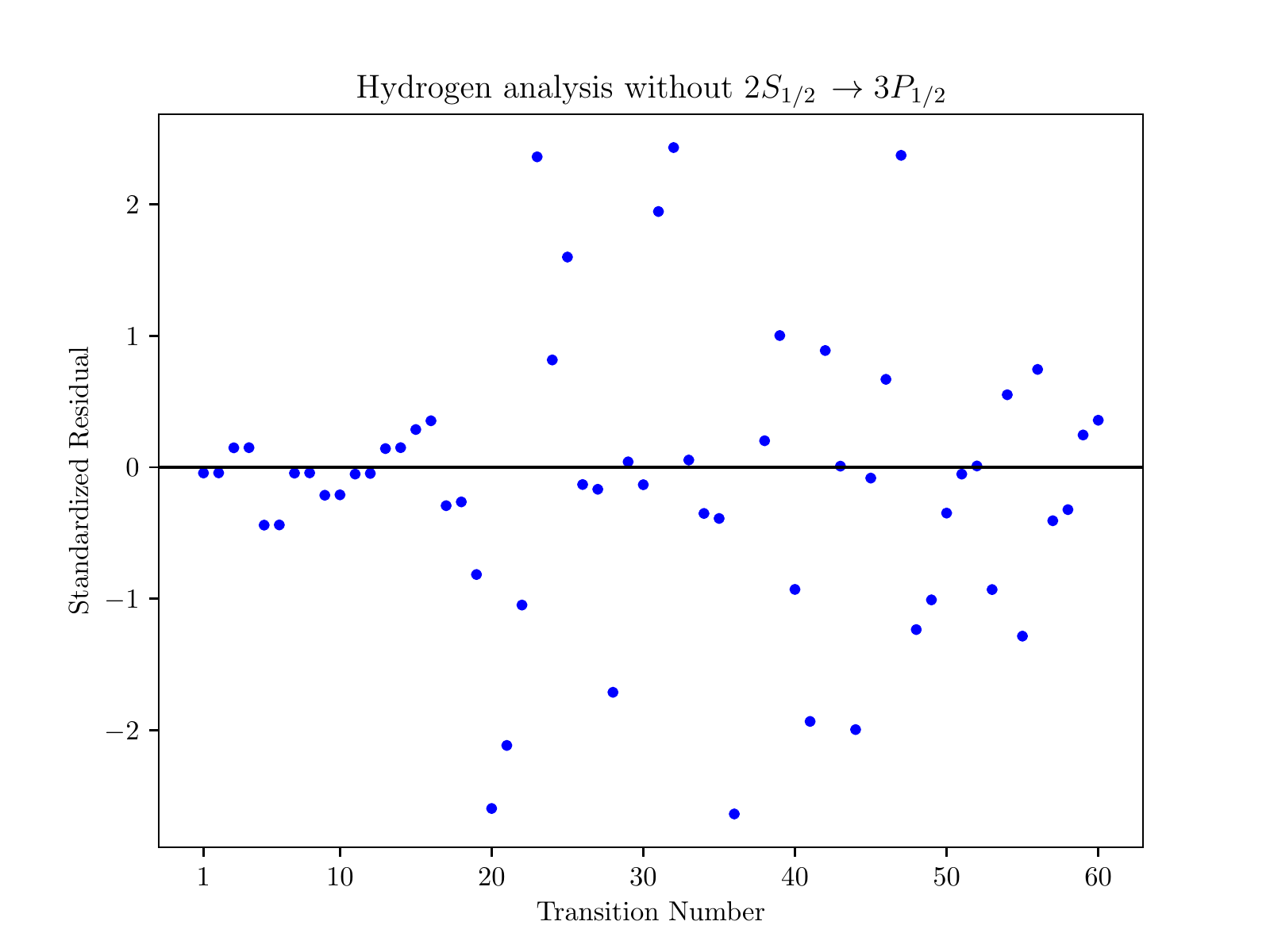}
  \end{center}
  \caption{Standardized residuals from the hydrogen analysis omitting the $2S_{1/2}\to3P_{1/2}$ transition. Transitions identified in Tables \ref{table:H_transitions_part1} and \ref{table:H_transitions_part2}.}
  \label{Hyd_firstcut_SPD_Std_Residuals_1-59}
  \end{figure}

\end{widetext}
\end{centering}

\ssec{Hydrogen analysis with multiple cuts}\label{lastcut_H_SPD_analysis}

We continue making cuts of one to two transitions until the standardized residuals indicate no substantially outlying data. In total, this requires cutting the transitions $1S_{1/2}\to3P_{3/2}$, $1S_{1/2}\to3P_{1/2}$, $1S_{1/2}\to2P_{1/2}$, $2S_{1/2}\to3P_{3/2}$, $2S_{1/2}\to3P_{1/2}$, $3S_{1/2}\to3P_{3/2}$, and $4P_{1/2}\to4D_{3/2}$, which corresponds to transitions 20, 21, 23, 36, 37, 44, and 47 in Tables \ref{table:H_transitions_part1} and \ref{table:H_transitions_part2}. AICc differences are given in Tables \ref{table:H_SPD_finalcut_OD=0} though \ref{table:H_SPD_finalcut_OD=2}, choosing RR211 as the best model. Although the minimum AICc value suggests that RR411 is naively the best, the principle of parsimony and an expectation that \emph{over-fitting} is possible suggests that the model with the local AICc minimum closest to RR000 should be chosen as the best, which is RR211. The parameter fits for RR211 are given in Table \ref{table:Hydrogen_final_SPD_RR211} and some of the AICc differences are given in Tables \ref{table:H_SPD_finalcut_OD=0}, \ref{table:H_SPD_finalcut_OD=1}, and \ref{table:H_SPD_finalcut_OD=2}.  The Akaike weights and some parameter values are given in Table \ref {table:A-weight-FS-EIH_fina_cut_HSPD} and the standardized residuals are shown in Figure \ref{Hyd_final_SPD_Std_Residuals_1-53}.

\begin{table}[H]
\centering
\caption{Fit parameters for hydrogen with RR211.}
\begin{tabular}{c | c}  \hline\hline
$\de_{(0) 0\f{1}{2}}$ & $2.53392(7)\times10^{-5}$\\  
$\de_{(2) 0\f{1}{2}}$ & $7.9(4)\times10^{-8}$\\  
$\de_{(4) 0\f{1}{2}}$ & $-7.0(9)\times10^{-8}$\\  
$\de_{(0) 1\f{1}{2}}$ & $2.66381(6)\times10^{-5}$\\  
$\de_{(2) 1\f{1}{2}}$ & $1.3(1)\times10^{-8}$\\  
$\de_{(0) 1\f{3}{2}}$ & $1.32944(6)\times10^{-5}$\\  
$\de_{(2) 1\f{3}{2}}$ & $1.2(1)\times10^{-8}$\\  
$\de_{(0) 2\f{3}{2}}$ & $1.33157(8)\times10^{-5}$\\  
$\de_{(2) 2\f{3}{2}}$ & $2(1)\times10^{-8}$\\  
$\de_{(0) 2\f{5}{2}}$ & $8.8684(8)\times10^{-6}$\\  
$\de_{(2) 2\f{5}{2}}$ & $1.4(4)\times10^{-8}$\\  
\hline
$E_0\,\[\text{eV}\]$ & $-13.598\,434\,599\,682(23)$\\  
$\a^{-1}$  & $137.035\,999\,186(19)$\\  \hline\hline
\end{tabular}
\label{table:Hydrogen_final_SPD_RR211}
\end{table}

\begin{table}[H]
\centering
\caption{Second-order Akaike information criterion differences, $\Delta_i$, for the $S,P,$ and $D$-channel analysis in hydrogen with multiple transition cuts. Here $O_D=0$. }
\begin{tabular}{c  | c c c c c }  \hline\hline
&&&$O_{P}$\\
$O_{S}$ & 0& 1 & 2 & 3 & 4\\ \hline
0&404.8 & 307.3 & 312.4 & 320.2 & 329.0 \\
1& 84.14 & 49.22 & 54.19 & 62.00 & 71.15 \\
2& 87.01 & 7.167 & 11.62 & 19.21 & 28.72 \\
 3&89.81 & 10.88 & 15.80 & 23.92 & 34.04 \\
 4&92.23 & 8.552 & 13.56 & 21.56 & 32.30 \\
 \hline\hline
\end{tabular}
\label{table:H_SPD_finalcut_OD=0}
\end{table}

\begin{table}[H]
\centering
\caption{Second-order Akaike information criterion differences, $\Delta_i$, as in Table \ref{table:H_SPD_finalcut_OD=0}.  Here $O_D=1$. Bolded values identify the models used for averaging.}
\begin{tabular}{c  | c c c c c }  \hline\hline
&&&$O_{P}$\\
$O_{S}$ & 0& 1 & 2 & 3 & 4\\ \hline
0&410.7 & 290.2 & 294.0 & 302.8 & 312.6 \\
1& 65.00 & 45.38 & 48.43 & 57.09 & 67.39 \\
2&68.20 & {\bf 0} & {\bf 3.490} & 11.74 & 22.40 \\
3& 71.32 & {\bf 2.542} & {\bf 5.411} & 14.20 & 25.55 \\
4& 73.39 & -1.831 & 0.4025 & 8.641 & 20.67 \\
 \hline\hline
\end{tabular}
\label{table:H_SPD_finalcut_OD=1}
\end{table}

\begin{table}[H]
\centering
\caption{Second-order Akaike information criterion differences, $\Delta_i$, as in Table \ref{table:H_SPD_finalcut_OD=0}.  Here $O_D=2$. Bolded values identify the models used for averaging.}
\begin{tabular}{c  | c c c c c }  \hline\hline
&&&$O_{P}$\\
$O_{S}$ & 0& 1 & 2 & 3 & 4\\ \hline
0& 416.0 & 294.2 & 298.5 & 308.4 & 319.5 \\
1& 53.42 & 25.63 & 23.03 & 32.28 & 43.79 \\
2& 54.16 & {\bf 3.172} & {\bf 9.333} & 18.88 & 31.02 \\
3& 47.69 & {\bf 7.764} & {\bf 14.31} & 24.72 & 37.75 \\
4& 48.56 & 3.154 & 9.869 & 19.61 & 33.46 \\
 \hline\hline
\end{tabular}
\label{table:H_SPD_finalcut_OD=2}
\end{table}

\begin{table}[H]
\centering
\caption{Akaike weights and parameters from the hydrogen $S,P,$ and $D$-channel analysis with multiple transition cuts.}
\begin{tabular}{c  |  c c c}  \hline\hline
Model & $w_i$ & $\a^{-1}$ & $E_I^{\text{(H)}} \text{eV}$ \\ \hline
RR211 & $0.5690$ &  $137.035\,999\,186(19)$ & $13.598\,434\,599\,682(23)$  \\  
RR221 & $0.0994$ &  $137.035\,999\,171(22)$ & $13.598\,434\,599\,697(25)$  \\  
RR311 & $0.1596$ &  $137.035\,999\,180(20)$ & $13.598\,434\,599\,683(23)$  \\  
RR321 & $0.0380$ &  $137.035\,999\,162(23)$ & $13.598\,434\,599\,701(25)$  \\  
RR212 & $0.1165$ &  $137.035\,999\,206(24)$ & $13.598\,434\,599\,682(25)$  \\  
RR222 & $0.0053$ &  $137.035\,999\,197(34)$ & $13.598\,434\,599\,690(31)$  \\  
RR312 & $0.0117$ &  $137.035\,999\,214(31)$ & $13.598\,434\,599\,681(25)$  \\  
RR322 & $0.0004$ &  $137.035\,9984(14)$ & $13.598\,434\,600\,10(77)$  \\  \hline\hline
\end{tabular}
\label{table:A-weight-FS-EIH_fina_cut_HSPD}
\end{table}

\begin{centering}
\begin{widetext}

\begin{figure}[H]
  \begin{center}
    \includegraphics[scale=.8]{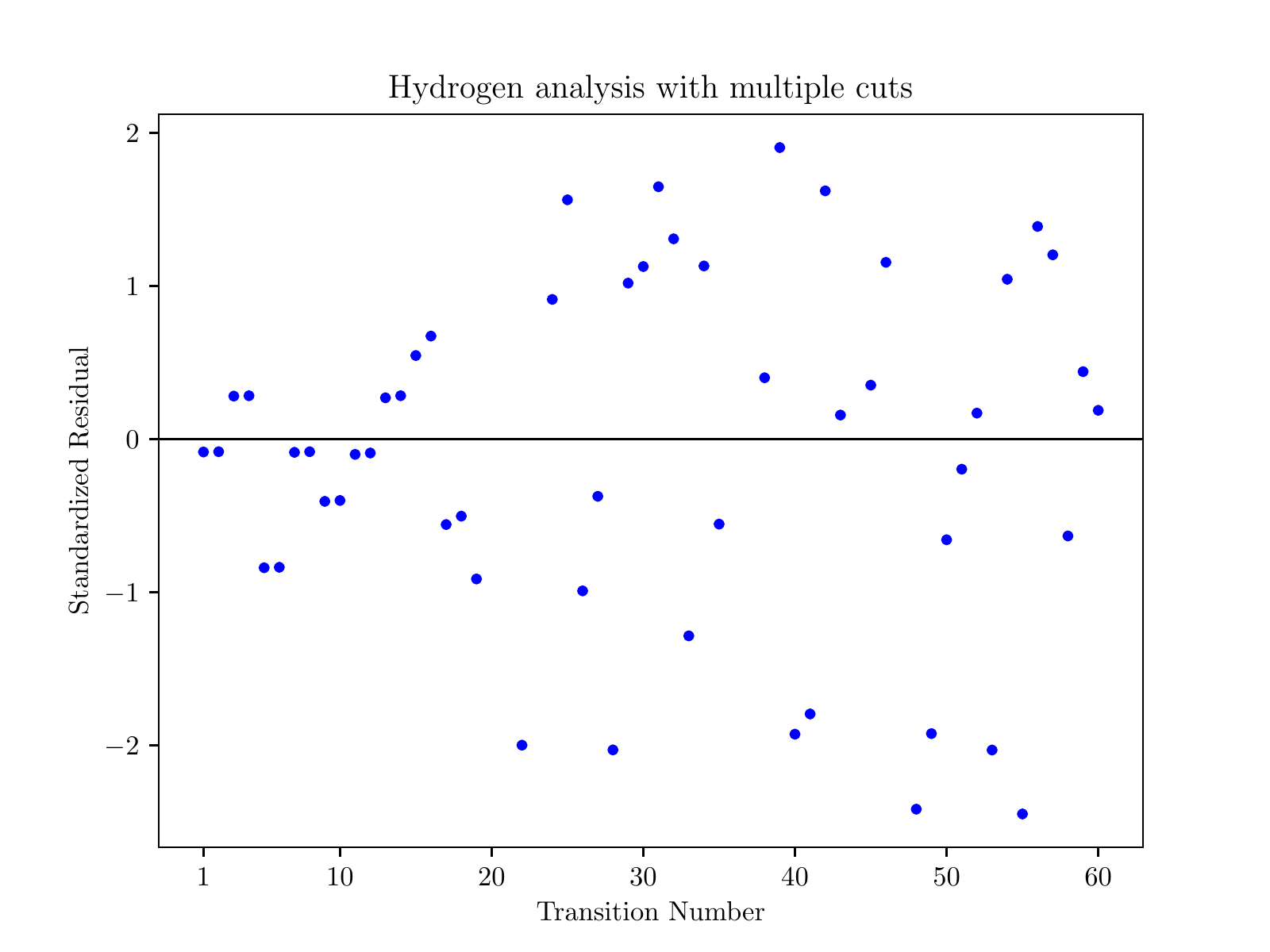}
  \end{center}
  \caption{Standardized residuals from the hydrogen analysis omitting several transitions; see text for details. Transitions identified in Tables \ref{table:H_transitions_part1} and \ref{table:H_transitions_part2}.}
  \label{Hyd_final_SPD_Std_Residuals_1-53}
  \end{figure}

\end{widetext}
\end{centering}

Here we repeat the analysis of Section \ref{Uncut_H_SPD_analysis} and \ref{firstcut_H_SPD_analysis_no_2S3P}. We determine
\begin{eqnarray}\label{alpha_inv_last_H_SPD}
\a^{-1} &=&  137.035\,999\,185 (23)_\text{data} (11)_\text{mass}\notag\\
&=&  137.035\,999\,185 (25)\,,
\end{eqnarray}
which is in agreement with the values in Table \ref{table:alpha_inv_determinations} at the level of $1.8\times10^{-10}$, with the exception of Ref. \cite{Parker:2018vye}; as described in Section \ref{amu_variation} below, the discrepancy here can be attributed to a discrepant determination of the atomic mass unit, $u$.  At the same time, we find
\begin{equation}\label{EHI_last_H_SPD}
E_I^{\text{(H)}} = 13.598\,434\,599\,684(25)\,\text{eV}\,,
\end{equation}
which is in agreement with the Standard Model prediction \eqref{QED_E0H} at the level of $1.8\times10^{-12}$. As part of this agreement, additional confirmation is possible for recent determinations of the proton radius, of particular interest given the so-called proton radius puzzle \cite{ubachs2020crisis}. Consider that the leading order nuclear size correction to the hydrogen ground state energy is
\begin{equation}
\Delta E_\text{NS}\(r_p\)=\f{8\pi^2}{3}\a^4 m_\text{red}^3 r_p^2 = 6.5\times 10^{-9}\,\text{eV}~\(\f{r_p}{1\,\text{fm}}\)^2\,,
\end{equation}
whereas the next-to-leading order correction is about $10^{-4}$ times smaller \cite{yerokhin2019theory}. This means that a variation in the proton radius would vary the ground state energy by approximately
\begin{equation}
\de\Delta E_\text{NS}\(r_p\) \simeq1.1\times 10^{-10}\,\text{eV}~\(\f{r_p}{0.84\,\text{fm}}\)\(\f{\de r_p}{0.01\,\text{fm}}\)\,,
\end{equation}
where $r_p\simeq 0.84\,\text{fm}$ is the currently accepted value \cite{Tiesinga:2021myr}. An error of $\de r_p=0.04\, \text{fm}$, corresponding to the older (incorrect) value of  $r_p\simeq 0.88\,\text{fm}$, would have likely resulted in disagreement between the Standard Model value \eqref{QED_E0H} and the data-driven value in \eqref{EHI_last_H_SPD}. This suggests a path toward further testing of nuclear structure models.

Finally, we can make contact with other analyses of hydrogen transition data by using this analysis to make a BSQED-independent determination of the Rydberg constant,
\begin{equation}
R_\infty =\frac{1}{2}\f{m_e c}{h}\a^2\,.
\end{equation}
To find its uncertainty we must consider the effect of the statistical uncertainty (due to fitting) as well as the uncertainty propagated from the mass-related input parameters. The latter are found to be sub-dominant to the former, which amount to $3.7\times10^{-3}\,\text{m}^{-1}$. Therefore, this analysis gives
\begin{equation}
R_\infty  =10\,973\,731.5718(37) \,\text{m}^{-1}\,.
\end{equation}
which is in agreement at the level of $3.4\times10^{-10}$ with the value as determined using BSQED analyses of specific atomic transitions \cite{Tiesinga:2021myr}.

\ssec{Predictions for omitted transitions}\label{sec:predictions}

Using the fits from Section \ref{lastcut_H_SPD_analysis}, here we predict values for the 7 transitions omitted from that analysis.  Predictions for a particular transition can be written in the form
\begin{equation}
\Delta \nu = f({\bf \Theta})\,,
\end{equation}
where $f$ represents $1/h$ times the right hand side of equation \eqref{transition_E_eqn} and ${\bf \Theta}$ is the vector of best-fit parameters from a particular fit. The uncertainty in $\Delta \nu$, $\sigma_{\Delta \nu}$ is given by
\begin{equation}\label{variance_estimate}
\sigma_{\Delta \nu}^2 = {\bf g}^T {\cal C} {\bf g}
\end{equation}
where the components of the vector ${\bf g}$ are
\begin{equation}
{\bf g}_a= \f{\d f}{\d {\bf \Theta}_a }
\end{equation}
and the covariance matrix has components
\begin{equation}
{\cal C}_{ab} = \text{cov}\({\bf \Theta}_a,{\bf \Theta}_b\)\,,
\end{equation}
both of which can be easily obtained with software, such as \emph{Mathematica}.

As an illustration, the main points of the analysis using the RR211 model on the transition $2S_{1/2}\to 3P_{1/2}$ are summarized here. With the  best-fit parameters for that model, given in Table \ref{table:Hydrogen_final_SPD_RR211}, equation \eqref{transition_E_eqn} predicts
\begin{equation}
\Delta \nu (2S_{1/2}\to 3P_{1/2})_\text{RR211} = 456\,681\,550\,679\,\text{kHz} \,.
\end{equation}
We take the components of ${\bf g}$ to be listed in the same order as that of Table \ref{table:Hydrogen_final_SPD_RR211}. Taking the derivatives of $f({\bf \Theta})_\text{RR211}$ with respect to those parameters and evaluating with the best-fit values,
\begin{equation}
{\bf g}=
\begin{pmatrix}
27\,420\\
6\,855\\
1\,714\\
-8\,124\\
-903\\
0 \\
0 \\
0 \\
0 \\
0 \\
0 \\
0 \\
-222\\
\end{pmatrix}
\text{kHz}\,,
\end{equation}
where each zero indicates that a parameter has no effect on the prediction for this particular transition.

 The covariance matrix output by the fitting software is omitted here for brevity. The result for the error estimate using \eqref{variance_estimate} is 
\begin{equation}
\(\sigma_{\Delta \nu}\)_\text{RR211}=14 \,\text{kHz}\,.
\end{equation}
The same procedure is followed for the remaining 7 models; a summary of results is given in Table \ref{table:2S_to_3P12_example}. A weighted average may be computed along with an unconditional error estimate, determined following the procedure described in Section \ref{Exploratory_Hyd_SD_Analysis}.

\begin{centering}

\begin{table}[H]
\centering
\caption{Akaike weights and predictions for the $2S_{1/2} \to 3P_{1/2}$ transition. }
\begin{tabular}{c  |  c c}  \hline\hline
Model & $w_i$ & $\Delta \nu$ [kHz] \\ \hline
RR211 & $0.5690$ &  $456\,681\,550\,679(14)$\\
RR221 & $0.0994$ &  $456\,681\,550\,659(20)$\\
RR311 & $0.1596$ &  $456\,681\,550\,682(14)$\\
RR321 & $0.0380$ &  $456\,681\,550\,659(20)$\\
RR212 & $0.1165$ &  $456\,681\,550\,665(16)$\\
RR222 & $0.0053$ &  $456\,681\,550\,659(20)$\\
RR312 & $0.0117$ &  $456\,681\,550\,660(20)$\\
RR322 & $0.0004$ &  $456\,681\,550\,659(20)$\\  \hline\hline
\end{tabular}
\label{table:2S_to_3P12_example}
\end{table}

\end{centering}

The final model-averaged prediction is
\begin{equation}\label{RR_2S3P12_pred}
\Delta \nu (2S_{1/2}\to 3P_{1/2})_\text{RR} = 456\,681\,550\,675(17)\,\text{kHz} \,.
\end{equation}
which should be compared with the ASD 2010 value
\begin{equation}\label{2S_3P12_ASD}
\Delta \nu (2S_{1/2}\to 3P_{1/2})_\text{ASD} = 456\,681\,549\,89(28)\,\text{kHz} \,,
\end{equation}
which originates in the measurement of Ref. \cite{zhao1986remeasurement}. The measured value in \eqref{2S_3P12_ASD} is smaller than the relativistic Ritz prediction \eqref{RR_2S3P12_pred} by approximately $2.8\sigma$, yet \eqref{RR_2S3P12_pred} is more than an order of magnitude more precise than the measurement. A new measurement of this transition frequency would be an excellent test of the overall approach presented here. The complete set of predictions for all 7 of the omitted transitions may be found in Table \ref{table:H_transitions_pred}.

\newpage
\begin{widetext}
\begin{centering}

\begin{table}[H]
\centering
\caption{Predictions of omitted transition frequencies in the hydrogen $S,P,$ and $D$-channel analysis. The last column is the non-relativistic Ritz interpolation using equation \eqref{nonrel_Ryd_Ritz}.}
\begin{tabular}{c  |  c c c}  \hline\hline
Transition &  Rel. Ritz [GHz] & ASD Measured [GHz] & ASD NR Ritz [GHz] \\ \hline
$\Delta \nu (2S_{1/2}\to 3P_{1/2})$  &  $456\,681.550\,675(17)$ & $456\,681.549\,89(28)$ & $456\,681.549\,96(21)$ \\  
$\Delta \nu (2S_{1/2}\to 3P_{3/2})$  &  $456\,684.800\,756(60)$ & $456\,684.800\,11(28)$  & $456\,684.800\,04(21)$\\  
$\Delta \nu (1S_{1/2}\to 3P_{1/2})$  &  $2\,922\,742.963\,862(17)$  & $2\,922\,728.6(8.6)$& $2\,922\,742.9632(2)$  \\  
$\Delta \nu (1S_{1/2}\to 3P_{3/2})$ & $2\,922\,746.213\,943(60)$ & $2\,922\,728.6(8.6)$ & $2\,922\,746.2132(2)$\\
$\Delta \nu (1S_{1/2}\to 2P_{1/2})$ & $2\,466\,060.355\,3394(37)$ & $2\,466\,068.0(4.1)$ & $2\,466\,060.355\,3404(81)$\\
$\Delta \nu (3S_{1/2}\to 3P_{3/2})$ & $2.935\,277(60)$ & $2.9334(12)$ & $2.935\,268(60)$\\
$\Delta \nu (4P_{1/2}\to 4D_{3/2})$ & $1.3689(13)$ & $1.3711(12)$ & $1.369\,10(25)$\\
 \hline\hline
\end{tabular}
\label{table:H_transitions_pred}
\end{table}
\end{centering}
\end{widetext}

\ssec{Effect of variation in the atomic mass unit}\label{amu_variation}

There is by now a known discrepancy in the determined value of the fine-structure constant from two different groups, \cite{Parker:2018vye} and \cite{Morel:2020dww}, based on their measurements of the absolute masses of rubidium and cesium, respectively. Consider then, swapping the role played by the rubidium values in Table \ref{table:mass-related_parameters} for those given in Table \ref{table:mass-related_parameters_Cs}.

\begin{table}[H]
\centering
\caption{Mass-related parameters}
\begin{tabular}{c c c}  \hline\hline
Quantity& Value & Reference$$\\ \hline
$h/m(^{133}\text{Cs})$ & $3.002\,369\,4721(12)\times10^{-9}\text{\,m$^2\,$s}^{-1}$ & \cite{Parker:2018vye}\\  
$m(^{133}\text{Cs})$ & $86.909\,180\,53\,10(60) \,u$ & \cite{huang2021ame}\\ 
  \hline\hline
\end{tabular}
\label{table:mass-related_parameters_Cs}
\end{table}
Here we repeat the analysis of Section \ref{lastcut_H_SPD_analysis}, performing the averaging and uncertainty analysis yet again. We determine
\begin{eqnarray}
\a^{-1} &=&  137.035\,999\,025 (23)_\text{data} (28)_\text{mass}\notag\\
&=&  137.035\,999\,025 (36)\,,
\end{eqnarray}
which is in agreement with Ref. \cite{Parker:2018vye} but not with any other determination in Table \ref{table:alpha_inv_determinations}. At the same time, we find
\begin{equation}
E_I^{\text{(H)}} = 13.598\,434\,599\,684(24)\,\text{eV}\,,
\end{equation}
which is still in agreement with the Standard Model prediction \eqref{QED_E0H} at the level of $1.8\times10^{-12}$. This may be pertinent because improved determinations of $E_I^{\text{(H)}}$ should be possible with new and/or improved spectroscopic data even without improvement in the determination of the amu or, likewise, a consensus between the authors of \cite{Parker:2018vye} and \cite{Morel:2020dww} on its value.

\ssec{Hydrogen analysis with fixed $\a$}

Here we fix the value $\a^{-1}=137.035\,999\,166(15)$, obtained recently through an improved measurement of the electron g-factor \cite{fan2022measurement} and repeat the analysis of the 7 transition frequency cuts described in Section \ref{lastcut_H_SPD_analysis}. The Akaike weights and ionization energies are reported in Table \ref{table:A-weight-FS-EIH_fixed_alpha_HSPD}.

\begin{centering}

\begin{table}[H]
\centering
\caption{Akaike weights and ionization energy fits for the hydrogen $S,P,$ and $D$-channel analysis with fixed $\a$ from Ref. \cite{fan2022measurement}.}
\begin{tabular}{c  |  c c c}  \hline\hline
Model & $w_i$  & $E_I^{\text{(H)}} \text{eV}$ \\ \hline
RR211 & $0.4632$  & $13.598\,434\,599\,7033(91)$  \\  
RR221 & $0.1968$  & $13.598\,434\,599\,7023(88)$  \\  
RR311 & $0.2028$  & $13.598\,434\,599\,6981(98)$  \\  
RR321 & $0.0878$  & $13.598\,434\,599\,6969(94)$  \\  
RR212 & $0.0348$  & $13.598\,434\,599\,718(14)$  \\  
RR222 & $0.0080$  & $13.598\,434\,599\,715(13)$  \\  
RR312 & $0.0054$  & $13.598\,434\,599\,710(17)$  \\  
RR322 & $0.0012$  & $13.598\,434\,599\,706(17)$  \\  \hline\hline
\end{tabular}
\label{table:A-weight-FS-EIH_fixed_alpha_HSPD}
\end{table}

\end{centering}

The measurement error in $\a^{-1}$ leads to a variation in the energy of $1.5\times10^{-11}$ eV. Altogether, 
\begin{eqnarray}
E_I^{\text{(H)}} &=& 13.598\,434\,599\,702(10)_\text{data}(15)_\alpha\,\text{eV}\notag\\
 &=& 13.598\,434\,599\,702(18)\,\text{eV}\,,
\end{eqnarray}
which is a marginal improvement to \eqref{EHI_last_H_SPD}, the value obtained in the variable-$\alpha$ analysis, and in agreement with the Standard Model prediction \eqref{QED_E0H} at the level of $1.3\times10^{-12}$.

\sec{Atomic deuterium}\label{Deuterium_Analysis}

\subsection{Deuterium with variable $\a$}\label{Initial_Deut_SPD_Analysis}

Here we analyize the 36 measured deuterium transitions from the NIST ASD 2010 hydrogen compilation \cite{NIST_ASD}, which includes all transitions between the 5 angular momentum channels $S_{1/2},P_{1/2},P_{3/2},D_{3/2}$, and $D_{5/2}$. These data are summarized in Table \ref{table:D_transitions}.  The fitting follows as in previous sections, and in particular involves fitting for the  fine-structure constant, $\a$. The Akaike weights and some parameter values are given in Table \ref{table:A-weight-FS-EID_initial_DSPD}.

\begin{centering}

\begin{table}[p!]
\centering
\caption{Deuterium transitions from Ref.  \cite{NIST_ASD}.}
\begin{tabular}{c  |  c c c}
Number& Transition &  Frequency (GHz)  \\\hline
1 & $ 1 \text{S}_{\text{1/2}}\to 2 \text{P}_{\text{3/2}} $ & $ 2\,466\,742.2(1.0) $ \\
 2 & $ 1 \text{S}_{\text{1/2}}\to 2 \text{S}_{\text{1/2}} $ & $ 2\,466\,732.407\,521\,70(14) $ \\
 3 & $ 2 \text{S}_{\text{1/2}}\to 12 \text{D}_{\text{5/2}} $ & $ 799\,409.184\,9668(64) $ \\
 4 & $ 2 \text{S}_{\text{1/2}}\to 12 \text{D}_{\text{3/2}} $ & $ 799\,409.168\,0372(85) $ \\
 5 & $ 2 \text{S}_{\text{1/2}}\to 10 \text{D}_{\text{5/2}} $ & $ 789\,359.610\,240(39) $ \\
 6 & $ 2 \text{P}_{\text{1/2}}\to 9 \text{D}_{\text{3/2}} $ & $ 781\,645.75(31) $ \\
 7 & $ 2 \text{P}_{\text{3/2}}\to 9 \text{D}_{\text{5/2}} $ & $ 781\,634.79(31) $ \\
 8 & $ 2 \text{P}_{\text{1/2}}\to 8 \text{D}_{\text{3/2}} $ & $ 770\,860.36(22) $ \\
 9 & $ 2 \text{S}_{\text{1/2}}\to 8 \text{D}_{\text{5/2}} $ & $ 770\,859.252\,8501(59) $ \\
 10 & $ 2 \text{S}_{\text{1/2}}\to 8 \text{D}_{\text{3/2}} $ & $ 770\,859.195\,7016(59) $ \\
 11 & $ 2 \text{S}_{\text{1/2}}\to 8 \text{S}_{\text{1/2}} $ & $ 770\,859.041\,2472(79) $ \\
 12 & $ 2 \text{P}_{\text{3/2}}\to 8 \text{D}_{\text{5/2}} $ & $ 770\,849.56(22) $ \\
 13 & $ 2 \text{P}_{\text{1/2}}\to 7 \text{D}_{\text{3/2}} $ & $ 755\,128.601(57) $ \\
 14 & $ 2 \text{P}_{\text{3/2}}\to 7 \text{D}_{\text{5/2}} $ & $ 755\,117.702(57) $ \\
 15 & $ 2 \text{P}_{\text{1/2}}\to 6 \text{D}_{\text{3/2}} $ & $ 730\,890.310(53) $ \\
 16 & $ 2 \text{P}_{\text{3/2}}\to 6 \text{D}_{\text{5/2}} $ & $ 730\,879.477(71) $ \\
 17 & $ 2 \text{P}_{\text{1/2}}\to 5 \text{D}_{\text{3/2}} $ & $ 690\,691.823(48) $ \\
 18 & $ 2 \text{P}_{\text{3/2}}\to 5 \text{D}_{\text{5/2}} $ & $ 690\,681.098(64) $\\
 19 & $ 2 \text{P}_{\text{1/2}}\to 4 \text{D}_{\text{3/2}} $ & $ 616\,690.175(38) $ \\
 20 & $ 2 \text{S}_{\text{1/2}}\to 4 \text{D}_{\text{5/2}} $ & $ 616\,689.59\,6718(37) $ \\
 21 & $ 2 \text{S}_{\text{1/2}}\to 4 \text{P}_{\text{3/2}} $ & $ 616\,689.141\,73(16) $ \\
 22 & $ 2 \text{S}_{\text{1/2}}\to 4 \text{S}_{\text{1/2}} $ & $ 616\,687.903\,573(20) $ \\
 23 & $ 2 \text{S}_{\text{1/2}}\to 4 \text{P}_{\text{1/2}} $ & $ 616\,687.769\,99(19) $ \\
 24 & $ 2 \text{P}_{\text{3/2}}\to 4 \text{D}_{\text{5/2}} $ & $ 616\,679.760(51) $ \\
 25 & $ 2 \text{P}_{\text{1/2}}\to 3 \text{D}_{\text{3/2}} $ & $ 456\,810.1143(21) $ \\
 26 & $ 2 \text{S}_{\text{1/2}}\to 3 \text{P}_{\text{3/2}} $ & $ 456\,809.062\,66(28) $ \\
 27 & $ 2 \text{S}_{\text{1/2}}\to 3 \text{P}_{\text{1/2}} $ & $ 456\,805.811\,72(28) $\\
 28 & $ 2 \text{P}_{\text{3/2}}\to 3 \text{D}_{\text{5/2}} $ & $ 456\,800.2259(16) $\\
 29 & $ 2 \text{P}_{\text{3/2}}\to 3 \text{S}_{\text{1/2}} $ & $ 456\,796.254(28) $\\
 30 & $ 2 \text{S}_{\text{1/2}}\to 2 \text{P}_{\text{3/2}} $ & $ 9.912\,59(29) $ \\
 31 & $ 3 \text{P}_{\text{1/2}}\to 3 \text{P}_{\text{3/2}} $ & $ 3.2508(11) $ \\
 32 & $ 3 \text{S}_{\text{1/2}}\to 3 \text{P}_{\text{3/2}} $ & $ 2.9351(49) $ \\
 33 & $ 4 \text{P}_{\text{1/2}}\to 4 \text{P}_{\text{3/2}} $ & $ 1.371\,80(31) $ \\
 34 & $ 2 \text{P}_{\text{1/2}}\to 2 \text{S}_{\text{1/2}} $ & $ 1.059\,281(60) $\\
 35 & $ 3 \text{P}_{\text{1/2}}\to 3 \text{S}_{\text{1/2}} $ & $ 0.315\,31(40) $\\
 36 & $ 4 \text{P}_{\text{1/2}}\to 4 \text{S}_{\text{1/2}} $ & $ 0.1332(59) $ \\
 \hline \hline
\end{tabular}
\label{table:D_transitions}
\end{table}
\end{centering}

\begin{centering}
~

\begin{table}[H]
\centering
\caption{Akaike weights and some parameter fits for the deuterium $S,P,$ and $D$-channel variable-$\alpha$ analysis. }
\begin{tabular}{c  |  c c c}  \hline\hline
Model & $w_i$ & $\a^{-1}$ & $E_I^{\text{(D)}} \text{eV}$ \\ \hline
RR100 & $0.7095$ &  $137.035\,999\,306(29)$ & $13.602\,134\,636\,459(39)$  \\  
RR110 & $0.0235$ &  $137.035\,999\,293(33)$ & $13.602\,134\,636\,459(44)$  \\  
RR200 & $0.1986$ &  $137.035\,999\,292(31)$ & $13.602\,134\,636\,459(40)$  \\  
RR210 & $0.0064$ &  $137.035\,999\,274(36)$ & $13.602\,134\,636\,459(45)$  \\  
RR101 & $0.0544$ &  $137.035\,999\,327(33)$ & $13.602\,134\,636\,459(40)$  \\  
RR111 & $0.0006$ &  $137.035\,999\,318(42)$ & $13.602\,134\,636\,459(47)$  \\  
RR201 & $0.0057$ &  $137.035\,999\,299(58)$ & $13.602\,134\,636\,459(45)$  \\  
RR211 & $0.0013$ &  $137.035\,999\,01(14)$ & $13.602\,134\,636\,459(88)$    \\  \hline\hline
\end{tabular}
\label{table:A-weight-FS-EID_initial_DSPD}
\end{table}
\end{centering}

\begin{centering}
\begin{widetext}
~
\begin{figure}[H]
  \begin{center}
    \includegraphics[scale=.8]{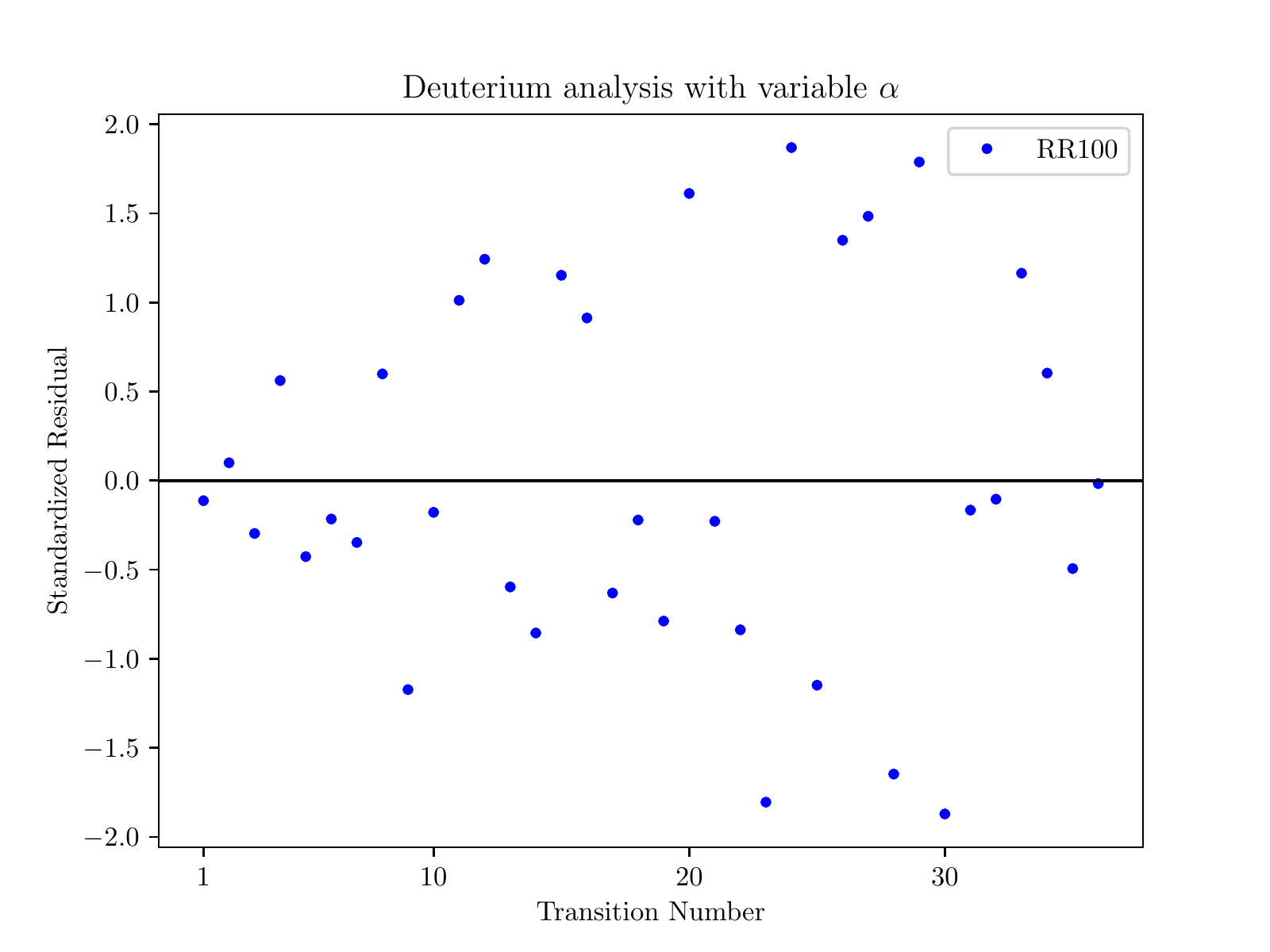}
  \end{center}
  \caption{Standardized residuals from the deuterium analysis performed with variable $\alpha$. Transitions identified in Table \ref{table:D_transitions}.}
  \label{Deut_uncut_SPD_Std_Residuals_1-36}
  \end{figure}
\end{widetext}
\end{centering}

From this initial  analysis we determine
\begin{eqnarray}\label{exploratory_D_alpha_inv_final}
\a^{-1} &=&  137.035\,999\,303 (31)_\text{data} (11)_\text{mass}\notag\\
&=&  137.035\,999\,303 (33) 
\end{eqnarray}
and
\begin{equation}\label{exploratory_E_I_D}
E_I^{\text{(D)}} = 13.602\,134\,636\,461(40)\,\text{eV}\,,
\end{equation}
The inverse fine-structure constant is far higher than what would be expected based on the values given in Table \ref{table:alpha_inv_determinations}, but this is consistent with a preference here for models with a low number of parameters, as discussed in Section \ref{Exploratory_Hyd_SD_Analysis}.

Standardized residuals are given in Figure \ref{Deut_uncut_SPD_Std_Residuals_1-36}. There is no obvious outlier data to consider cutting. It appears that, given the relative paucity of deuterium transition data (as compared to hydrogen), it is not possible to make a reliable determination of both the fine-structure constant and deuterium ionization energy.

\ssec{Deuterium with fixed $\a$}

In this section we impose a value of the fine structure constant determined in Ref. \cite{fan2022measurement},
\begin{eqnarray}
\a^{-1} &=&   137.035\,999\,166 (15)\,,
\end{eqnarray}
and repeat the analysis from the previous section. The Akaike weights and ionization energies are given in Table \ref{table:A-weight-FS-EID_final_DSPD}.

\begin{centering}

\begin{table}[H]
\centering
\caption{Akaike weights and ionization energies for the deuterium $S,P,$ and $D$-channel analysis with fixed $\a$. }
\begin{tabular}{c  |  c c c}  \hline\hline
Model & $w_i$  & $E_I^{\text{(D)}} \text{eV}$ \\ \hline
RR211 & $0.7240$  & $13.602\,134\,636\,539(25)$  \\  
RR201 & $0.1649$  & $13.602\,134\,636\,541(29)$  \\  
RR311 & $0.0455$  & $13.602\,134\,636\,537(26)$  \\  
RR301 & $0.0140$  & $13.602\,134\,636\,542(30)$  \\  
RR210 & $0.0211$  & $13.602\,134\,636\,612(19)$  \\  
RR200 & $0.0250$  & $13.602\,134\,636\,613(21)$  \\  
RR310 & $0.0018$  & $13.602\,134\,636\,612(20)$  \\  
RR300 & $0.0037$  & $13.602\,134\,636\,614(21)$  \\  \hline\hline
\end{tabular}
\label{table:A-weight-FS-EID_final_DSPD}
\end{table}

\end{centering}

The measurement error in $\a^{-1}$ leads to a variation in the ionization energy equal to $9.3\times10^{-12}$ eV, whereas a variation due to mass-input uncertainties, determined by a Monte Carlo simulation as in Section \ref{Exploratory_Hyd_SD_Analysis}, amounts to  $6.8\times10^{-12}$ eV. Altogether, 
\begin{eqnarray}\label{final_E_I_D}
E_I^{\text{(D)}} &=& 13.602\,134\,636\,5430 (283)_\text{data} (93)_\alpha (68)_\text{mass} \,\text{eV}\notag\\
&=& 13.602\,134\,636\,543 (31) \,\text{eV}
\end{eqnarray}
which agrees with the Standard Model prediction at the level of $2.3\times10^{-12}$.



\section{Summary and Discussion}\label{SummaryAndDiscussion}

This article presents a first attempt at fitting experimental atomic transition frequencies with the relativistic Ritz approach, a semi-empirical long-distance effective theory of hydrogen-like atoms. The quantum defect is written as a series expansion in low energy whose series coefficients parameterize physical effects whose range is shorter than the (infinite) range of the Coulomb interaction. Such effects include radiative QED corrections, as well as nuclear structure effects and any short-ranged beyond-the-standard-model dynamics.

Atomic hydrogen transition data was analyzed first in an exploratory way, considering various quantum defect truncations,  or model choices. The best analysis yielded a value for the fine-structure constant, $\alpha=1/137.035\,999\,185 (25)$, in agreement with the determination using other methods and without using bound-state QED. Simultaneously, the ionization energy  was found to be $E_I^{(\text{H})}=13.598\,434\,599\,684(25)\,\text{eV}$, in agreement with the Standard Model prediction at  $1.8\times10^{-12}$. Some outlying transition data was discovered and, through continued cutting of outliers and re-analysis, I conclude that at least seven transitions are worth excluding from analysis. Predictions were made for these transitions which may be verified in future experiments.

As for atomic deuterium, there was insufficient data to simultaneously determine both the fine-structure constant and ionization energy. Using a fixed value of $\alpha$ coming from a recent electron g-factor measurement, the ionization energy was determined to be $E_I^{(\text{D})}=13.602\,134\,636\,543 (31) \,\text{eV}$, in agreement with the Standard Model at the level of $2.3\times10^{-12}$.

Historically, in order to measure $\a$ using spectroscopic data, bound-state QED has been employed to extract the Rydberg constant from which the determination of $\a$ is made. The nuclear radius also must be accounted for with that approach, but here such detailed modeling is not needed. The relativistic Ritz approach instead leverages the statistical power in using a modest number of atomic transitions, some of which have only been measured with modest precision. This may complement other approaches that apply bound-state QED to a smaller number of highly precise measurements. Another feature of the relativistic Ritz approach is that it can be used even if short-ranged beyond-the-standard-model phenomena are discovered to affect hydrogen-like atoms.

New and more precise measurements of atomic transitions, especially within hydrogen, deuterium, and positronium, may provide more accurate determinations of the fine-structure constant and ground state energies of simple atoms, thus allowing for a new way to test the Standard Model with precision atomic physics. Future work should determine what precision may be possible with future experimental data, but this is beyond the scope of the present article.

\section*{Acknowledgments}
I am grateful to Alexander Kramida for correspondence and for conversations with Gabe Duden, Maggie Rasmussen, Gloria Clausen, Fr\'ed\'eric Merkt, and Dylan Yost.  This research was funded, in part, though a Charles A. Dana Research Fellowship at Norwich University.

\bibliography{scibib}

\end{document}